\documentclass[aps,prd,preprint,nofootinbib]{revtex4-1}
\usepackage{amsmath, bm, natbib, latexsym}
 \begin{document}

 \title{Plane wave holonomies in loop quantum gravity I: symmetries and gauges}

\author{Donald E. Neville}

\email{dneville@temple.edu}

\affiliation{Department of Physics,
Temple University,
Philadelphia 19122, Pa.}


\newcommand{\rA}{\mathrm{A}}
\newcommand{\rC}{\mathrm{C}}
\newcommand{\rD}{\mathrm{D}}
\newcommand{\rE}{\mathrm{E}}
\newcommand{\rF}{\mathrm{F}}
\newcommand{\rH}{\mathrm{H}}
\newcommand{\rK}{\mathrm{K}}
\newcommand{\rL}{\mathrm{L}}
\newcommand{\rM}{\mathrm{M}}
\newcommand{\rN}{\mathrm{N}}
\newcommand{\rS}{\mathrm{S}}
\newcommand{\rU}{\mathrm{U}}
\newcommand{\rV}{\mathrm{V}}
\newcommand{\rY}{\mathrm{Y}}
\newcommand{\rmd}{\mathrm{d}}
\newcommand{\rmi}{\mathrm{i}}

\newcommand{\E}[2]{\mbox{$\mathrm{E}^{#1}_{#2}$}}
\newcommand{\Eav}{\mbox{$\bar{\mathrm{E}}^z_Z$}}
\newcommand{\A}[2]{\mbox{${\mathrm{A}}^{#1}_{#2}$}}
\newcommand{\Nt}{\mbox{\underline{{N}} }}
\newcommand{\Ht}{\mbox{$\tilde{\mathrm{H}} $} }
\newcommand{\Etwo}{\mbox{$^{(2)}\tilde{\mathrm{E}} $}\ }
\newcommand{\Etld }{\mbox{$\tilde{\mathrm{E}}  $}\ }
\newcommand{\Vtwosq}{\mbox{$(^{(2)}{\mathrm{V} })^2 $}}
\newcommand{\Vtwo}{\mbox{$^{(2)}{\mathrm{V}} $}}
\newcommand{\Ethree}{\mbox{$^{(3)}{\mathrm{E}} $}}
\newcommand{\etwo}{\mbox{$^{(2)}{\mathrm{e}} $}}
\newcommand{\ethree}{\mbox{$^{(3)}{\mathrm{e}} $}}


\newcommand{\nn}{\nonumber \\}
\newcommand{\rta}{\mbox{$\rightarrow$}}
\newcommand{\rla}{\mbox{$\leftrightarrow$}}
\newcommand{\eq}[1]{equation (\ref{#1})}
\newcommand{\Eq}[1]{Equation (\ref{#1})}
\newcommand{\eqs}[2]{equations (\ref{#1}) and (\ref{#2})}
\newcommand{\Eqs}[2]{Equations (\ref{#1}) and (\ref{#2})}
\newcommand{\bra}[1]{\langle #1 \mid}
\newcommand{\ket}[1]{\mid #1 \rangle}
\newcommand{\braket}[2]{\langle #1 \mid #2 \rangle}
\newcommand{\cd}{\delta_{(c)}\,}
\newcommand{\td}{\tilde{\delta}\,}
\newcommand{\si}{\mbox{sgn}}

\begin{abstract}

This is the first of two papers which study the
behavior of the SU(2) holonomies of loop quantum gravity (LQG),
when they are acted upon by a unidirectional, plane gravity wave.  Initially,
the LQG flux-holonomy variables are treated as classical,
commuting functions rather than quantized operators, in a limit where variation
from vertex to vertex are small and fields are weak.  Despite the
weakness of the fields, the field equations are not linear.
Also, the theory can be quantized, and the expectation values
of the quantum operators behave like their classical analogs.
Exact LQG theories may be either local or non-local.  The present paper
argues that a wide class of non-local theories
share non-local
features which survive to the semiclassical limit, and
these non-local features are included
in the classical limit studied here.
An appendix
computes the surface term required when the propagation direction
is the real line rather than $\mathrm{S}_1$.  Paper II
introduces coherent states, constructs a damped sine wave
solution to the constraints,  and solves for the behavior
of the holonomies in the presence of the wave.

\end{abstract}

\maketitle
\section{Introduction}

    This and a succeeding paper \cite{SineWave} investigate the
behavior of loop quantum gravity (LQG) fluxes and holonomies in the
presence of a gravitational plane wave. The behavior of
traditional, metric variables in presence of a weak gravitational
wave is well known. Metric waves are discussed in most classical textbooks;
quantization of the theory is straightforward.
However, no corresponding discussion has been given for the
flux-holonomy variables characteristic of LQG.

   Most calculations in this paper are  classical.
Constraints may be imposed either
at the classical level, or
at the quantum level in the Dirac manner.
However, the unidirectional constraints are second-class
and \emph{must} be treated at the classical level.  This
leads to Dirac
brackets, which are messy.  Computation of  Dirac brackets for the most
general classical theory (two polarizations, no gauges fixed)
is especially complicated.  We choose to fix gauges and impose symmetries
at the classical level, which leads to
the simplest possible Dirac brackets.
Quantization (early in paper II) is then straightforward,
requiring a single paragraph:
replace (Dirac) brackets by commutators; choose
factor orderings.

    After that one-paragraph foray into the quantum theory,
the paper reverts to the classical side.  The Hilbert space is
based on coherent states.
These states turn quantum operators back into classical, commuting
functions.  Consider a quantum constraint which is a product
of operators  $O_1\, O_2 \cdots$, acting on a coherent state
$\ket{\mbox{coh}}$.
\begin{equation}
    ( O_1\, O_2 \cdots)\ket{\mbox{coh}} \cong (O_1(cl)  O_2(cl) \cdots)\ket{\mbox{coh}}.
\label{Quantum=Classical}
\end{equation}
Coherent states are peaked at specific values of flux
and holonomy, and the operators become functions $O_i(cl)$
evaluated at those peak values.

    Paper II constructs a sinusoidal solution to
the classical constraints.  One then reads \eq{Quantum=Classical} right to
left: in a regime where coherent states are applicable,
a classical solution to the constraints implies the vanishing of the quantum constraint.

    The classical results
carry over to the quantum theory.  The expectation value of a
quantum operator varies with the plane wave in the same way as the
corresponding classical variable.

    A classical LQG theory should
possess the following six features.
The basic variables are holonomies and fluxes; they have support only on
a lattice; variables are invariant under spatial diffeomorphisms;
areas and volumes are quantized; the
field theory limit (the limit where the lattice disappears)
is correct; and the theory is adequately regulated.
("Adequately":  1/volume  need not be regulated.
Since spins are large in the classical limit, the
volume does not vanish.)

    (The term "classical" is slightly ambiguous when applied to an
LQG theory.  "Classical" may refer to a theory using commuting
flux-connection variables, with support on a continuum.  Or "classical" may
mean a theory using commuting flux-holonomy variables, with support only on a lattice.
Papers I-II use the second meaning of classical.  The theory defined
on the continuum will be referred to as classical field theory, or
simply field theory (FT).

    We  would assert it is not necessary to derive the classical
theory as a limit of an exact LQG theory.  If we did so, the
paper would turn into a test of that exact theory, rather than
an investigation of holonomies.  If the exact theory
did not have one or more of the six features listed above, one
would reject the exact theory as inadequate.  I.~e. one starts from the
above six properties, then tests the exact theory by requiring
it possess these properties in the limit.  The six properties
are a more basic starting point than any exact theory.

   Every \emph{exact} theory should also possess the above
six features (excepting the comment about the volume).
Two further assumptions are introduced to make the theory
easier to solve, as well as closer to the classical limit: dynamical
quantities vary slowly from vertex to
vertex; and holonomies are small.  Precise definitions
of "slowly" and "small" are given in sections \ref{SS}
and \ref{SV}.   These assumptions produce a theory which
is less non-linear than exact LQG theories, but still non-linear.
The constraints are not quadratic in the variables.  Despite the
weak field assumption, the theory is unlike
the linear weak field theory of geometrodynamics.

    For works which do construct classical limits
 starting from exact theories, see Han \cite{Han},
and Giesel and Thiemann \cite{Giesel}.
Han uses a path integral/spin foam approach, rather than the
canonical approach used here.  Giesel and Thiemann take the
semiclassical limit of the Master Constraint, rather than
the usual scalar and vector constraints, S and V.  However,
the semiclassical Master Constraint is essentially a sum of squares,
$\rS^2 + \rV^a \rV_a$, so that the solutions of the
follow-on paper (which are annihilated by S and V) should also
be solutions to the Master Constraint.

    Both the above treatments are
quite general; there is no discussion of the planar case
or gauge-fixing. Bannerjee and Date \cite{BannDate} construct
an exact LQG theory which is  specialized to the planar case;
see also Hinterleitner and Major \cite{HintMaj}.
Both papers use a Bohr quantization of the
transverse degrees of freedom.  There are no holonomies
along transverse directions (x,y), only holonomies along
the longitudinal direction (propagation direction, z).
Transverse degrees of freedom are represented by two scalars
(essentially, the magnitudes of the axes of the polarization
ellipse) plus an angle (non-zero, if the ellipse axes do not coincide
with the x and y axes).  If the goal is a study of the behavior of
holonomies in the presence of a gravitational wave, then the theory
must use holonomies along x and y, rather than Bohr quantization.

    In addition, both
those theories are local.  I.~e., ~holonomic loops used to define
the field strengths at the nth vertex remain infinitesimally
close to that vertex.  A non-local, "nearest neighbor",  theory
uses loops which include the nearest neighbor vertices at n $\pm$ 1.
Motivation for including non-local features is given at the
beginning of section \ref{Nonlocal}.

    The present work has both a primary and a secondary
goal. The primary goal is to study the behavior
of holonomies in the presence of a gravitational wave.
The secondary goal is to assume  the exact theory is non-local, and
study the effect of non-local features on the theory.

    In deciding which non-local features to include, again,
one should not start from one exact non-local theory.
Rather, one should identify
non-local features which are common to a large class
of non-local theories,and survive to the classical
limit.
Consider the class of theories which treat both nearest neighbors
on an equal footing.  I.~e., let the vertices along z be
indexed by integers $n_z$.  If the model includes a non-local holonomic
loop starting at $n_z$ and going to $n_z + 1$, then it must also include a loop
starting at $n_z$ and going to $n_z-1$, and with equal weight.  Since there is no
reason to favor one nearest neighbor over the other, this class of
theories is likely to be large.

     Given equal treatment of both nearest neighbors,
which non-local features survive?  Section \ref{Nonlocal} takes
the semiclassical limit of a sample non-local model which treats both nearest
neighbors equally.   Let $h_a(n_z)$ be a holonomy along transverse edges
a = x or y (direction of propagation = z), located at
vertex numbered $n_z$.  A local
definition for the derivative of $h_a$ would be
\[
   [ h_a(n_z + \epsilon/2) - h_a(n_z -\epsilon/2)]/\epsilon.
\]
 $\epsilon$ is an infinitesimal regulator which cancels out
at the end of the calculation.  The non-local model of section
\ref{Nonlocal}, when taken to the limit of small connections
and slow variation,
replaces the above local definition by a nearest-neighbor, non-local
generalization.
\begin{equation}
    [ h_a(n_z + 1) - h_a(n_z - 1)]/2 \Delta z.
\label{NLDifference}
\end{equation}
$\Delta z$, the distance between vertices, must be
small, like $\epsilon$, so that the above ratio
is a good approximation to the derivative.
Unlike $\epsilon$, $\Delta z$  is not taken to zero at
the end of the calculation.  In a
non-local approach the difference, and
not the derivative, is fundamental.

    Similarly, the model uses a non-local version
of $h_z$, the holonomy along the propagation direction.
\begin{eqnarray}
    h_z(n_z) &=& [h_z(n_z,n_z+ \epsilon) + h_z(n_z-\epsilon, n_z)]/2 \quad \mbox{(local)};\nn
    h_z(n_z) &=& [h_z(n_z,n_z+1) + h_z(n_z-1,n_z)]/2  \quad \mbox{(non-local)}; \nn
    h_z(a,b) &:=& \exp [i \int_a^b \rA^Z_z \sigma_Z/2].
\label{NLHolonomy}
\end{eqnarray}

    The model considered in section \ref{Nonlocal} is
relatively simple. Appendix \ref{nonlocal} discusses a more complex
non-local model, which also treats nearest
neighbors equally.  This model uses non-standard grasps, as
well as definitions of differences and z holonomies
which do not agree with the non-local definitions given at
\eqs{NLDifference}{NLHolonomy}.  Nevertheless, in the semiclassical
limit the non-local definitions in the exact model are
replaced by the definitions at \eqs{NLDifference}{NLHolonomy}.
The two models offer strong support for the idea that a non-local
model which treat neighbors equally will always possess a semiclassical limit
with differences and holonomies given by \eqs{NLDifference}{NLHolonomy}.

    Note the above non-local modifications of difference
and holonomy are certainly plausible.  Even before studying any
exact non-local model, if one wished to treat nearest
neighbors equally, then
the above central difference and z holonomy
are certainly the simplest possibilities.  Presumably the reader
could skip the detailed study of the non-local models
on a first reading.

    Section \ref{SSH} constructs a classical Euclidean Hamiltonian.
Section \ref{SSK} constructs LQG extrinsic curvatures.
Section \ref{Gauss} constructs the  Gauss constraint and
the Lorentzian Hamiltonian.  Sections \ref{SinglePol}, \ref{DiffGauge},
and \ref{Unidirectional} discuss single polarization,
diffeomorphism, and unidirectional constraints respectively.
Section \ref{Nt} discusses boundary conditions
at infinity.

    Experimentally, it is clear that SU(2) holonomies
(which are just rotation matrices) can be superimposed to form a coherent
state, because the earth (for example) presumably is described
by a superposition of Legendre polynomials (rotation
matrices again); yet both its angular momentum and
conjugate angle are sharp.

    Theoretically, however, matters are less clear.
Coherent states eventually spread, unless the system has
the equally spaced energies characteristic of the SHO.  It is
necessary to show under what conditions the spreading is
limited.  This is done in the succeeding paper which
introduces the coherent states.

    The term "planar" is a slight misnomer: the theory
does not have full planar symmetry in the xy plane.
With suitable choice
of coordinates, the Killing vectors become $\partial/\partial x,
\partial /\partial y$, implying that all functions are
independent of x and y.
However, this is translational invariance, not full planar
symmetry, which would require isotropy with respect to rotations
in the xy plane.
Isotropy is inconsistent with the presence of waves.  Vibrations of the
usual cloud of test particles are described by an ellipse,
which picks out preferred directions.  The translational
invariance implies the ellipse is the same everywhere in the
xy plane.

    For a quantization of plane waves using  geometrodynamics
variables, see Mena Marug\'{a}n and Montejo \cite{Montejo}.

\subsection{Conventions}

    Throughout, indices from the middle of the alphabet
i, j, $\cdots$ range over coordinates x, y, z on the manifold.  Indices
from the beginning of the alphabet a, b, $\cdots$
range over x, y only, where z is the direction of propagation.
Similarly , indices I, J, K range over coordinates
X, Y, Z in the
local free-fall frame.  Indices A, B $\cdots$ range over transverse
directions X, Y only.

    When expanding 2 x 2 matrices, I use
Hermitean sigma matrices, rather than anti-Hermitean tau matrices.
A typical Lie group valued operator would be written
\begin{equation}
    \mathbf{O}_i := \mathrm{O}^I_i \,\mbox{{\boldmath $\sigma$}}_I
\label{sigmaNotS}
\end{equation}
The sigma matrices, and
bold face for matrices, will be suppressed except when it is necessary to
emphasize the matrix character of an equation.  It should be clear from context which
quantities are sigma-valued.  Usually the operator
in \eq{sigmaNotS} will be written simply as $\mathrm{O}_i$.

    In LQG densitized cotriads are written as
area two-forms,
\[
    \rE^i_I(n) \,dx^j \,\wedge \,dx^k \,\epsilon_{ijk}/2,
\]
and connections are written as one-forms, \A{J}{j} $dx^j$.  The area
and line integrals in the definitions of triad and connection
guarantee simpler transformation
properties under spatial diffeomorphisms; also, the
[holonomy, triad] commutator will contain enough integrations
to kill the delta function.  Usually, the
area and line integrals will be suppressed.  E.~g. $\mathrm{E}^i_I$
will be written as

\begin{equation}
    \mathrm{E}^i_I \, dx^j \wedge dx^k \epsilon_{ijk}/2! \,\rta \,\mathrm{E}^i_I.
\label{defE}
\end{equation}

\subsection{Initial gauge fixing}

    Because of the planar symmetry,
Husain and Smolin are able to choose gauges
which simplify the \Etld and connection
fields \cite{HusSmo}.  These choices reduce the
general, 3 + 1 dimensional case to the
planar case; they therefore precede all the
gauge choices to be made in this paper.
\begin{eqnarray}
    \mathrm{E}^a_Z &=& \mathrm{E}^z_A = 0; \nn
    \mathrm{A}^Z_a &=& \mathrm{A}^A_z = 0.
\label{HusSmoGauges}
\end{eqnarray}
a = x,y; A = X,Y.
These choices fix the SU(2) rotations around axes X,Y and
the diffeomorphisms in transverse directions x,y.  Three
constraints survive: the scalar constraint, the vector
constraint for z diffeomorphisms, and the Gauss constraint
for rotations around Z: SU(2) $\rta$ U(1).

    Since the only \A{I}{z} which survives has
I = Z,  holonomies
along the longitudinal z direction are quite simple,
involving only the rotation generator $\rS_Z$ for
rotations around Z.
\[
    \exp(i \int \, \mathrm{A}^Z_z \cdot \mathrm{S_Z}).
\]
Conversely, the
transverse holonomies (those along the x and y directions  of the
spin network) contain no $\mathrm{S}_Z$ and involve $\mathrm{S}_X,\mathrm{S}_Y$ only.

\subsection{Topology of the spin network}
\label{topology}

    As a convenience for readers not familiar with
the usual network used in the planar case, this section
includes a description of the topology.

    In the z direction (direction of propagation of the wave)
the spin network has the topology of the
real line.  The line includes a series of vertices,
labeled by integers $n_z$.  The vertices are connected by edges,
which may be labeled by their endpoints, as $(n_z,n_{z+1})$.

    In directions transverse to propagation, there are two possible
approaches.  The first approach is easiest to relate to
the full, three-dimensional case.  Give each vertex on
the original z axis three integer coordinates: ($n_x = 0,
n_y = 0, n_z$).  Construct a three
dimensional rectangular lattice by drawing a
congruence of lines, all parallel to the original z axis.
All lines have the identical arrangement of vertices, but
differ in their x and y coordinates,  $n_x = \pm 1,
\pm 2, \cdots, n_y = \pm 1, \pm 2,\cdots.$  Connect neighboring
vertices having the same $n_z$ with edges $(n_x,n_{x+1}),
(n_y, n_{y + 1})$.  In this way one fills out
a full, three dimensional rectangular lattice.

    Each member of the congruence is labeled
by a pair of indices $(n_x, n_y)$, and each vertex by a
triplet $(n_x, n_y, n_z)$.  Because of the translational invariance,
physics will be independent of $(n_x, n_y)$.  We will refer
to this as the "congruence" picture.  (This is a slight
abuse of notation, since members of a traditional congruence
are labeled by continuously varying parameters, rather
than discrete integers ($n_x,n_y$).)

    The second method for handling the transverse directions
is simpler topologically, but a little
harder to relate to the three dimensional case.  Construct a
small cubic box surrounding
each vertex.  Equip each face with an outward normal.
Call a face positive (negative) if its normal points in the
positive (negative) coordinate direction.  Consider the holonomy
with support on edge $(n_x, n_{x+1})$.  It leaves a cube at position
$n_x$, passing through the positive x face, then enters the
nearest neighbor cube at $n_{x+1}$, through a negative x face.  Because
of the planar symmetry, the holonomy entering the negative x
face of cube $n_{x+1}$ must be identical to the holonomy entering
the negative x face of cube $n_x$.  Therefore one could
give the edge $(n_x, n_{x+1})$ the topology of a circle:
the holonomy leaves cube $n_x$ through the positive x face, travels
along $(n_x,n_{x+1})$ (now a circle, rather than a straight line)
and reenters $n_x$ through the negative x face.

    The congruence
has now disappeared.  There is only a real line R in the z direction, and
two $\mathrm{S}_1$ edges leaving each vertex in the x and y directions.
We will refer to this as the "$\mathrm{S}_1$ picture". The
R $\times \mathrm{S}_1 \times \mathrm{S}_1$ topology is simpler for calculations: but for thinking,
it is perhaps better to use the congruence: one has more
assurance the results will generalize to three dimensions.

     In the congruence picture, it is natural to refer
to the smallest rectangular area enclosed by x and y edges
as an "xy plane".  We use this terminology, even though
in the $\rS_1$ picture this area has the topology of a
torus.  Similarly, an area bounded by two
neighboring edges in the z direction and two neighboring
x edges will be called the "xz plane".  In the $\rS_1$ picture this area
has the topology of a cylinder.

\section{Approximations}
\label{Approx}


    This section
proposes specific small field and slow variation assumptions.
These assumptions simplify calculations; they also bring the
theory close to the limit where quantum behavior goes over
to classical behavior.

\subsection{The small field (small sine) approximation}
\label{SS}

    One can obtain the field theory (FT) limit of LQG
(lattice $\rta$ continuum) by expanding
the holonomy as
\begin{equation}
        h_i = \exp(i\int \rA_i \cdot \rS)\cong 1 + i\int \rA_i \cdot \rS \quad \mbox{(FT)}.
\label{QFTApproximation}
\end{equation}
This expansion is too drastic for present purposes.
It replaces a bounded expression
by an unbounded one.  The following, small sine
approximation is less drastic, in that the
bounded expression is replaced by another bounded
expression, because the connection remains inside a
holonomy.  Expand the basic spin 1/2 holonomy
in sigma matrices:

\begin{eqnarray}
    \mathbf{h}_i &=& \exp(i \mbox{\boldmath $\sigma$} \cdot \hat{n}^{(i)} \,\theta_i/2) \nn
                    &=& \mathbf{1} \cos(\theta_i/2) +  i \sin (\theta_i/2) \,\hat{n}^{(i)}\cdot \mbox{\boldmath $\sigma$} .
\label{hSigmaExpansion}
\end{eqnarray}
$h_i$ is a rotation through $\theta$ around  an axis
given by $\hat{n}$.  Now expand the expression in powers of
sine, keeping out  to linear in sine.
\begin{equation}
    \mathbf{h}_i \cong \mathbf{1}  +  i \sin (\theta_i/2) \,\hat{n}_i\cdot \mbox{\boldmath $\sigma$}+\mbox{order} \sin^2(\theta_i/2). \quad \mbox{(SS)}
\label{SmallSineApprox}
\end{equation}
SS stands for for small sine.  The function on the right remains
bounded.

    When carrying out this expansion, it is a little simpler
 to write each holonomy as
 \begin{eqnarray}
    h &:=& \bar{h} + \hat{h}; \nn
    \bar{h} &=& (h + h^{-1})/2 = \mathbf{1} \cos(\theta_i/2);\nn
    \hat{h} &=& (h - h^{-1})/2 = i \sin (\theta_i/2) \,\hat{n}^{(i)}\cdot \mbox{\boldmath $\sigma$}.
 \label{defhbarhhat}
 \end{eqnarray}
 Then
 \begin{equation}
    h \cong 1 + \hat{h}, \:\mbox{(SS)}
 \label{SSLimitHhat}
 \end{equation}
 which is a more compact notation
 not involving explicit factors of $\sin (\theta/2)$.  In this notation,
 the passage from small sine to field theory is (compare  \eqs{SSLimitHhat}{QFTApproximation})
 \begin{equation}
    -2i \,\hat{h}_{i} \rightarrow \int A_i \cdot \sigma  \:\mbox{(FT)}.
 \label{QFTLimitHhat}
 \end{equation}

    When expanding a given constraint in sines, how many terms
 should be kept?   When taking LQG to the field theory limit,
 one must keep terms out to order $\mathrm{A}^2$, in order to recover the
 usual field theory Hamiltonian.  Therefore, in the small sine expansion of the
 constraints, one must keep terms out to order $\hat{h}^2$ = order $\sin^2$.  This will guarantee that
 the small sine limit has the same FT limit as full LQG.

    The small sine replacement is simply a recognition that certain
 terms in the scalar constraint are negligible in the weak field
 limit.  \emph{SS need not be used everywhere in the theory}.  If
 a given constraint or a Hilbert space state is already tractable,
 in its exact form, there is no need to simplify further. In
 particular, the follow-on paper constructs a Hilbert space of
 states.  Those states are products of exact spin 1/2
 holonomies (no SS expansion).  The states are coherent, so that
 their behavior (when acted upon by holonomy or flux operators,
 including the sine) is already simple; a SS expansion of
 states would be pointless.

    Since the basic holonomy is just an SU(2) rotation
matrix, the products of holonomies at each vertex form
representations of SU(2). (The longitudinal holonomies along z
form representations of U(1).)  One might question the validity of the SS
approximation, because the
kinematic dot product based on SU(2) or U(1) Haar measure integrates
over all values of $\theta_i$, therefore over all values of
$\sin(\theta_i/2)$, not just small values.

    Here the coherent states come to the rescue.
Coherent states are designed to be peaked simultaneously
at both a coordinate and a conjugate momentum ($\theta_a$ and
typical spin $\mathrm{L}_a$, a = x,y; or $\theta_z$ and
typical z component of spin m).  If peak values of $\theta_i$ are
chosen small, then matrix elements
will be dominated by small values of $\sin (\theta_i/2)$, and
the small sine approximation will be valid.

    The wavefunctional can be peaked at small $\sin (\theta_i/2)$,
only if typical angular momenta $\mathrm{L}_a$ (and z components m)
are moderately large.  As is typical for coherent states, the  standard deviations of $\theta_a$
and its conjugate momentum $\mathrm{L}_a$
are inverses of each other.  The standard deviations are order $1/\sqrt{\mathrm{L}_a}$
and $\sqrt{\mathrm{L}_a}$ respectively.  Sharp
$\theta_a$ therefore requires moderately large $\mathrm{L}_a$, $1/\sqrt{\mathrm{L}_a}\ll \pi$.
The small sine approximation breaks down if the
representations of the rotation group occurring at a given vertex
have too small values of total angular momentum.

    The small sine assumption, discussed above, does not
explicitly mention large quantum numbers.  Nevertheless,
it is clear from the discussion of coherent states
that the small sine assumption will not work unless
quantum numbers are large.

\subsection{The slow variation assumption}
\label{SV}

    In the classical  limit one expects slow
variation of dynamical quantities
from one vertex to the next \cite{Bojowald}.  Slow variation
implies that a plot of the quantity versus vertex index
$n_z$ looks like a smooth curve, rather than a union of
piecewise smooth segments.

    To make this idea more precise, define
central and forward differences by
\begin{eqnarray}
    \delta_c f(n) &=& (f(n+1)-f(n-1))/2; \\
    \delta_f f(n) &=& f(n+1) - f(n).
\label{defFirstDifference}
\end{eqnarray}
The slow variation assumption is

\begin{equation}
    (\delta f/f) \ll 1,
\label{SlowVariation}
\end{equation}
where $\delta$ may be either difference.

    The slow variation assumption also applies to
higher differences.  Define second differences by
\begin{eqnarray}
    \delta_c^{(2)}f(n) &:=& [ \,\delta_c f(n+1) -\delta_c f(n-1) \,] \nn
        &=& [\,f(n+2) -2 f(n) + f(n-1) \,]/4;\nn
    \delta_f^{(2)}f(n-1) &:=& [ \,\delta_f f(n) -\delta_f f(n-1) \,] \nn
        &=& [ \,f(n+1) -2 f(n) + f(n-1) \,].
\label{defSecondDifference}
\end{eqnarray}
If $\delta f/f$ is negligible,  $(\delta f/f)( \delta g/g)$ is more so.
Let g = $\delta f$.
\begin{eqnarray}
    (\delta f/f)( \delta g/g) &=& (\delta f/f)( \delta (\delta f)/( \delta f) \nn
                & =&  \delta^{(2)} f/f \ll 1.
\label{SmallSecondDerivative}
\end{eqnarray}
The second difference is of second order in
small differences.

    The slow variation assumption may be thought
of as a consequence of the small sine assumption.
Sines contain time derivatives $\partial/\partial ct$,
since, from  \eq{QFTLimitHhat},
\[
    2\sin(\theta_i/2) = -2i \hat{h} \rightarrow \int A_i \cdot \sigma ,
\]
and A contains the exterior derivative K.
The differences correspond to space
derivatives $\partial/\partial z$.  Since the
excitations are massless, time and space
derivatives should be comparable; both are small
if one is small.  Since the two assumptions
are closely connected, for brevity sometimes
we will refer to small sine, slow variation
simply as small sine.

    Since the space derivatives are of the
same order as the sines, and we are keeping
out to order $\sin^2$, we must keep differences
out to order $(\delta E/ E)^2$.   The
spin connections $\Gamma^I_i$ are order
$\delta E/ E$, since they contain one derivative
and are homogeneous of degree zero in the
triads.  Therefore $ \Gamma$ terms must
be kept out to order $\Gamma^2$.

\section{Local \emph{vs.} non-local}
\label{Nonlocal}

    Initial formulations of LQG
used local field strengths \cite{QSD}.  Smolin
used the renormalization group to argue that a local
formulation does not allow
propagation of information from one vertex to the next
\cite{SmolLR,NevLR}.  In
response, Thiemann \cite{MasterConstraint,ThiemannText}
proposed his "master constraint" program,
which allows non-local field strengths, while preserving
a constraint algebra free of anomalies.  (The two issues,
anomalies and non-locality, are closely connected, because
the original, local formulation is anomaly free.)

    A large quantum number calculation is not suited for checking
the master constraint program, or equivalently checking for
the presence of anomalies.  In quantum geometrodynamics,
anomalies arise when a constraint commutator produces
a metric component to the right of a constraint.  The
corresponding result in LQG would be triad components
to the right of a constraint.  Triads have matrix elements
of order the spin of the state, i.~e. if a coherent state
has angular momentum peaked at L, the matrix element will
be order L.  If the triad is moved to the left of a
basic spin 1/2 holonomy in the constraint, the holonomy
will change the spin of the state by order unity, therefore
change the matrix element of the triad, in its new
position, by order unity.  The fractional change
in the matrix element, on moving the triad to the
left, is then order $\Delta L/L \sim 1/L$, which is
negligible in the limit of large quantum numbers.  (See also
the further comments on anomalies in section \ref{Nt}.)

\subsection{A non-local model}
\label{Exact}

    If a classical non-local model includes contributions from both nearest
neighbors, and includes them with equal weights, then the
weak field limit will involve central differences and
averaged z holonomies, as at \eqs{NLDifference}{NLHolonomy}.
The following, specific model shows how this happens.

    The model employs holonomic loops, nearest neighbor non-local.
 \begin{align}
   \rF_{xy}(n_z) &= 2i h_x(n_z)^{-1} h_y(n_z)^{-1} h_x(n_z) h_y(n_z)/\Delta x \Delta y;\nn
     \rF_{za}(n_z,n_z+1) &= 2i h_a(n_z) h_z(n_z,n_z+1)^{-1} h_a(n_z+1)^{-1} h_z(n_z,n_z+1)/\Delta x^a \Delta z;\nn
      \rF_{za}(n_z,n_z-1) &= -2i h_a(n_z) h_z(n_z-1,n_z) h_a(n_z-1)^{-1} h_z(n_z-1,n_z)^{-1}/\Delta x^a \Delta z.
 \end{align}
On each line, add the Hermitean conjugate.  The loop for $\rF_{ij}$ is a finite
rather than infinitesimal rectangle in plane ij.
The loops $\rF_{za}(n,n')$ run from vertex n to nearest neighbor vertex n$'$,
then return to vertex n.

    These field strengths may be taken to the small sine limit
by systematically replacing one or two
$h_i^{\pm 1}$ by $\pm \hat{h}_i$, then replacing the remaining $h_i$ by unity.
A $[\hat{h}_a(n_z), \hat{h}_a(n_z+1)]$ term in $\rF_{za}$ cancels because the
commutator of transverse sigma matrices can give only unity or $\sigma_z$,
and this  vanishes when $\rF_{za}$ is traced with
the triad-dependent factor in the Hamiltonian, $\Sigma^{za}$.  That factor contains
no $\sigma_Z$
\[
    \Sigma^{za} \sim [\rE^z_Z \sigma_Z, \rE^a_A \sigma_A]/|e| \sim \sigma_B,\quad \mbox{A, B $\neq$ Z}.
\]
The small sine limits are then
\begin{align}
    F_{xy}(n_z) &= [\hat{h}_x, \hat{h}_y]2i/\Delta x \Delta y; \nn
    F_{za}(n_z,n_z+1) &= (-2i)[\hat{h}_a(n_z+1) - \hat{h}_a(n_z)]/\Delta x^a \Delta z \nn
            & \qquad + (2i)[\hat{h}_z(n_z,n_z+1), \hat{h}_a(n_z+1)]/\Delta x^a \Delta z;\nn
     F_{za}(n_z,n_z-1) &= (-2i)[\hat{h}_a(n_z) - \hat{h}_a(n_z-1)]/\Delta x^a \Delta z \nn
            & \qquad + (2i)[\hat{h}_z(n_z-1,n_z), \hat{h}_a(n_z-1)]/\Delta x^a \Delta z. \quad \mbox{(SS)}
 \label{SSF}
 \end{align}
 The factors of (-2i) are needed for the FT limit; cf. \eq{QFTLimitHhat}.
 (Another factor of 2i is generated by the
 commutators of the sigma matrices.)  There is only one possible $\rF_{xy}$ because
 only holonomies at $n_z$ are available for its construction.  There
 are two $\rF_{za}$ because $n_z$ has two nearest neighbors.

    Now apply slow
variation to the model.  Both loops $\rF_{za}(n_z,n_z \pm 1)$ start
from the same vertex $n_z$ and are multiplied by the same
triad factor $\Sigma^{za}(n_z)$.  Nearest
neighbor contributions from both $n_z -1$ and $n_z + 1$ are included,
and with equal weights, because there
seems no reason to favor one nearest neighbor over the other.

    Treating the neighbors equally has significant consequences.
Consider first the forward difference terms.
\begin{eqnarray}
    H_e &=& \cdots + [\rF_{za}(n_z,n_z+1) + \rF_{za}(n_z,n_z-1)]\Sigma^{za}(n_z) \nn
        &\sim & \cdots + \hat{h}_a(n_z+1) - \hat{h}_a(n_z) + \hat{h}_a(n_z) - \hat{h}_a(n_z-1)\nn
        & = & \cdots + 2 \cd \hat{h}_a(n_z) .
\label{FwdDiffToCentDiff}
\end{eqnarray}
The two forward differences in $\rF_{za}(n_x, n_z \pm 1)$ have combined
into one central difference, \eq{defFirstDifference}.

    There are also consequences for the  commutator terms.
\begin{equation}
   \rH_e \sim  \cdots + (2i)\{[\hat{h}_z(n_z,n_z+1), \hat{h}_a(n_z+1)] + [\hat{h}_z(n_z-1,n_z), \hat{h}_a(n_z-1)]\}. \nn
\label{Commutators1}
\end{equation}
Each operator may be split up into an average plus a difference.
\begin{eqnarray}
     \hat{h}_z(n_z,n_z \pm 1) &=& [\hat{h}_z(n_z,n_z+1) +\hat{h}_z(n_z-1,n_z)]/2 \nn
                            & & \quad \pm [\hat{h}_z(n_z,n_z+1) - \hat{h}_z(n_z-1,n_z)]/2 \nn
                            &:= & \hat{h}_z(n_z) \pm   \td \hat{h}_z(n_z);\nn
     \hat{h}_a(n_z \pm 1) &\cong& \hat{h}_a(n_z) \pm \delta_f \hat{h}(n_z). \quad \mbox{(SV)}
\label{Commutator2}
\end{eqnarray}
The last line uses slow variation (SV).
From \eq{defSecondDifference}, the exact formula for $\hat{h}_a(n_z - 1)$ is
\[
    \hat{h}_a(n_z - 1) = \hat{h}_a(n_z) - \delta_f \hat{h}(n_z) +  \delta_f^{(2)}\hat{h}_a(n-1),
\]
The slow variation assumption was used to drop the second difference.
These expansions may be inserted into \eq{Commutators1} for the commutator.
\begin{equation}
    \rH_e \sim \sum_{\pm}(2i)[\hat{h}_z(n_z) \pm \td \hat{h}_z/2, \hat{h}_a(n_z) \pm \delta_f \hat{h}(n_z)] .
\end{equation}
Now expand in the small differences.  Because the sum is even under
(+ $\rla$ -), terms with an odd number of differences vanish.  The
leading term involves the average, $ \hat{h}_z$ times the \emph{local}
holonomy $\hat{h}_a(n_z)$; the $\hat{h}_a(n_z \pm 1)$ have disappeared.
The term linear in differences vanishes.
The term quadratic in differences is down by $(\delta f/f)^2$ and
may be dropped.

    We have now arrived at \eqs{NLDifference}{NLHolonomy}:
central rather than forward differences; and an averaged z holonomy.
This outcome is a consequence of the small sine, slow variation assumptions
and the decision to include both nearest neighbor
field strengths with equal weights.  A non-local model which weights
nearest neighbors equally
will yield a limit with
central differences, local xy holonomies, and averaged z holonomies.

    In this limit one could replace differences
by derivatives, because differences
approach derivatives when variation from
vertex to vertex is small.  However, if the non-local
approach has any validity, the future of LQG will
involve differences. It is therefore helpful to retain
some non-local features in the present calculation.
It is reassuring that use of differences and averaged
holonomies causes no
problems, at least at this SS, SV level.

\subsection{Brackets involving $\hat{h}_z$}

    The $\hat{h}_z(n_z)$ defined at \eq{NLHolonomy} is non-local:
it does not commute with \E{z}{Z}($n_z \pm 1).$
Assuming the h have the same
Poisson brackets as $\exp[ i \int A \cdot \sigma/2]$, the non-locality comes from
the basic bracket
\begin{equation}
    \{h_z(n_z,n_z+1), \rE^z_Z(n_z \mbox{ or }n_z+1)\} = i(\kappa\gamma/2)(\sigma_z/2) h_z(n_z,n_z+1).
\label{hEzCommutator}
\end{equation}
$\kappa$ = 8 $\pi$ G; $\gamma$ is the Immirzi parameter.
\begin{equation}
    \mathrm{A}_a^A := \gamma \mathrm{K}_a^A + \Gamma_a^A.
\label{defAgamma}
\end{equation}
There is a factor of 1/2 on the right in \eq{hEzCommutator} because the
grasps occur at endpoints of the integration ranges, therefore
integrals are over only half a delta function.

    Despite the non-locality, in practice $\hat{h}_z(n_z)$
commutes like a local variable.  Once again slow variation
comes to the rescue.  Typically, $h_z(n_z$) occurs in a sum
or is commuted with a sum.  For example,
\begin{align}
   \{h_z(n_z), \sum_m g(m) \rE^z_Z (m)\} &= i(\kappa\gamma/2)(\sigma_z/2)\{h_z(n_z,n_z+1)[g(n_z+1) + g(n_z)] \nn
                                   & \qquad + h_z(n_z-1,n_z)[g(n_z-1) + g(n_z)]\}/2 ;\nn
              g(n_z+1) + g(n_z)  &= 2 g(n_z) + \delta_f g(n_z);\nn
              g(n_z-1) + g(n_z) &= 2 g(n_z) - \delta_f g(n_z) + \delta_f^{(2)}g(n_z).
\label{hzCommutator}
\end{align}
After neglect of terms of order $(\delta f/f)^2$, the commutator collapses to
\[
    i(\kappa\gamma/2)(\sigma_z/2) h_z(n_z) g(n_z) \times 2, \;\mbox{(SV)}
\]
which is just the local result.

    The bracket, \eq{hzCommutator}, involves h rather than $\hat{h}$;
but one can extend the proof to $\hat{h}$ by using the basic \eq{hEzCommutator}.
\begin{eqnarray}
    \{\hat{h}(n_z,n_z \pm 1), \rE^z_Z (n_z) \} &=& \{[h - h^{-1}, \rE^z_Z\}/2 \nn
                &=& i(\kappa\gamma/2)(\sigma_z/2)[[h + h^{-1}]/2\nn
                &=& i(\kappa\gamma/2)(\sigma_z/2)\bar{h}.
\label{hatzCommutator}
\end{eqnarray}
If this commutator occurs in a context where it is multiplied by
a term of order sine, one can approximate the final $\bar{h}$
by unity.

\section{A small sine, non-local $\rH_e$}
\label{SSH}

        We now propose the following small sine
LQG Euclidean Hamiltonian.

\begin{eqnarray}
   - \mathrm{N H}_e &+ \mathrm{ST}& =   \sum_{nz} \mathrm{N}(n_z)\{ \mathrm{F}^Z_{xy}(n_z) \,\mathrm{E}^x_J \,\mathrm{E}^y_K \,\epsilon_{ZJK}\nn
    &                &\qquad  + \mathrm{F}^A_{za} \,\mathrm{E}^z_Z \,\mathrm{E}^a_B \,\epsilon_{AZB}\}/\kappa \mid e\mid) + \mathrm{ST}; \nonumber \\
    &\mathrm{F}_{xy}(n_z) &= \mathrm{F}^Z_{xy}(n_z) \,\sigma_Z = 2i [ \,\hat{h}_x(n_z),\hat{h}_y(n_z) \,];\nn
    &\mathrm{F}_{za}(n_z)&= \mathrm{F}^A_{za}(n_z) \,\sigma_A = 2 i[ \,\hat{h}_z(n_z),\hat{h}_a(n_z) \,] \nn
     &       & \quad + (-2i)\delta_c \hat{h}_a(n_z). \qquad \mbox{(SS, SV)}
\label{SSHe}
\end{eqnarray}

The field strengths are given by the leading, order ($\sin$ + $\sin^2$) terms in the
small sine approximation.  The exact theory is assumed to treat nearest neighbors
symmetrically, and the Hamiltonian is modeled after the weak field limits of the
nearest neighbor models considered in section \ref{Nonlocal} and \ref{nonlocal}.
Consequently, the above Hamiltonian  involves central differences, rather than derivatives
or forward differences.  Also, the exact $\rF_{za}$ may
contain $\hat{h}_a(n_z \pm 1)$, but the small sine
limit contains only $\hat{h}_a(n_z)$; and the z holonomy is replaced by the
non-local average given at \eq{NLHolonomy}.

    The triads
\begin{equation}
    \Sigma^{ijK} := \mathrm{E}^i_I \mathrm{E}^j_J \epsilon^{IJK}/|e| \;\mbox{(QFT)}
\label{QFTSigma}
\end{equation}
are moved to the right, a standard choice.  The triads are "double
grasp".  E.~g.~ triad $\mathrm{E}^z(n)$ has support on xy areas
on both the incoming and outgoing sides of vertex n, so that
$\mathrm{E}^z(n)$ grasps both the incoming and the outgoing
z holonomy at vertex n.  The model considered in appendix
\ref{nonlocal} employs triads which grasp only
incoming or only outgoing holonomies, but not both.  However, in the small sine
limit, the single grasp triads are replaced by double grasp triads.
 The volume $e$ need not be regulated in this limit, since
it does not vanish for large quantum numbers.

    ST stands for "surface term",  required because  the z axis is the real line
rather than $\mathrm{S}_1$ (as in the  Gowdy model).
Since the surface term gives the energy, in the
follow-on paper it will be possible to calculate the energy of the plane wave.
The surface term is calculated in Appendix \ref{surface}.

\section{Flux-holonomy algebra}

    The flux-holonomy algebra for
these quantities is determined by the assumption that the \Etld
grasp both incoming and outgoing holonomies at a given vertex.  In the transverse
case, the double area grasp produces an anticommutator.
\begin{eqnarray}
    \{\mathrm{E}^a_A, h_a \} &=&i (\gamma\kappa/2)[\sigma_A/2, h_a]_{+}; \nn
    \{\mathrm{E}^z_Z, h_z \} &=& i (\gamma\kappa/2) [\sigma_Z/2, h_z]_{+}.
\label{fluxalgebra}
\end{eqnarray}
The area integrations in the E's (suppressed) combine with
the line integration in the holonomy to cancel the delta
functions.  For comparison, second line
exhibits the grasp of the longitudinal holonomy. That
result agrees with \eq{hzCommutator}, because
\[
    [\sigma_Z/2, h_z]_{+} = (\sigma_Z/2) h_z \times 2.
\]
The algebra for the $\hat{h}$ was derived
at \eq{hatzCommutator}.
\begin{eqnarray}
    \{\mathrm{E}^a_A, \hat{h}\} &=& i (\gamma\kappa/2)[\sigma_A/2, \bar{h}]_{+} \nn
                & \cong & i (\gamma\kappa/2)\sigma_A \quad \mbox{(SS)}
\label{hathAlgebra}
\end{eqnarray}

\section{Small sine expression for the extrinsic curvature}
\label{SSK}

    Thiemann \cite{QSD} has proposed a two step process for constructing a
regulated extrinsic curvature.  His procedure uses the Poisson brackets
\begin{eqnarray}
    \{|e|, H_e\} &\sim& K \cdot E := K_i^I E^i_I; \nn
    h_i \{ h_i^{-1},\, K \cdot E\}  &\sim&  K_i; \nn
    |e| &=& \sqrt{ \si(e) \, \mbox{det} E}.
\label{ThiemannK}
\end{eqnarray}
sgn(e) is the sign of e and det E.
If one inserts the SS Hamiltonian \eq{SSHe} on line
one above, the result for extrinsic curvatures and spin connections is
\begin{eqnarray}
    \gamma \mathrm{K}_i &=& -2i \hat{h}_i - \Gamma_i ;\nn
    \Gamma^I_j \mathrm{E}^j_I &=& \si(e) \, (\cd \Sigma^{zmM}) \Sigma^{ni}_M \epsilon_{mni}/2! := \Gamma \cdot \mathrm{E}; \nn
    \Gamma^M_j \mathrm{E}^j_I &=& -\si(e) \, (\cd \Sigma^{zmM}) \Sigma^{ni}_I \epsilon_{mni}/2! \nn
                    && \qquad + \Gamma \cdot \mathrm{E} \, \delta^M_I.
\label{SSKGamma}
\end{eqnarray}
The $\mathrm{E}^j_J$ are left in place because
the $\Gamma_j$ in the Hamiltonian
typically occur contracted with a triad.

    The functions $\Gamma = \Gamma [\Etld]$ in \eq{SSKGamma} are
identical to the classical ones, except z derivatives of the \Etld
are replaced by central differences.  To understand how this happens,
start from the SS $\rH_e$.  It
has terms linear and quadratic in the $\hat{h}$.  The
 \{$\mid e \mid, \rH_e$\} bracket removes one $\hat{h}$ (compare
 \eq{hathAlgebra}). $\rK \cdot \rE$ then contains
 terms linear in the $\hat{h}$ and terms independent of
 $\hat{h}$.  Since the subsequent bracket with h does not remove
 an $\hat{h}$, the $\Gamma$ come from the
 terms in  $\rK \cdot \rE$ independent of $\hat{h}$;
 equivalently, they come from the terms in $\rH_e$ linear in $\hat{h}$.

    The terms in $\rH_e$ linear in $\hat{h}$ are the
 central difference terms.
 \begin{eqnarray}
    \rH_e + \mathrm{ST} &\sim& \cd \hat{h}_a \, \Sigma^{za} + \cdots + \mathrm{ST}\nn
        &=& - \hat{h}_a \, \cd  \Sigma^{za} + \cdots + \mbox{no ST}.
 \label{IBP1}
 \end{eqnarray}
 The last line, the difference analog of an
 integration by parts, produces a surface term
 which is canceled by ST.  It also produces
 the $\cd \Sigma$ terms present in \eq{SSKGamma}.
 The additional $\Sigma$ factor (the factor with no $\cd$) comes from the
 bracket of $\mid e \mid$ with the $\hat{h}_a$ in \eq{IBP1}. $\Box$

     Proof that an "integration by parts" maneuver is
possible when dealing with differences rather than derivatives: start from
the following formula,  which is exactly true.
\begin{eqnarray}
      \cd (A\Sigma)(n) =& \cd A(n) (1/2)[ \,\Sigma(n+1) + \Sigma(n-1) \,] \nn
                    &\quad +(1/2)[ \,A(n+1) + A(n-1) \,]\cd \Sigma(n).
\label{ExactDistributive}
\end{eqnarray}
Compare this to the distributive formula for derivatives.  \Eq{ExactDistributive}
contains averages such as  (1/2)[ A(n+1) + A(n-1) ],
where the distributive law for derivatives has just A(n).
From \eq{defSecondDifference} the sum A(n+1) + A(n-1) equals
2 A(n) plus a forward second derivative.  The
slow variation assumption, \eq{SmallSecondDerivative}, can be used to drop the
second derivative.  Then
\begin{equation}
    \cd (A\Sigma)(n) = \cd A(n) \,\Sigma(n) + A(n)\,\cd \Sigma(n). \quad \mbox{(SV)}
\label{ApproximateDistribution}
\end{equation}
This formula more closely resembles the corresponding relation
for derivatives.  If the $ A(n)\,\cd \Sigma(n)$ term is moved
to the left-hand side  of the equation, the result is an
integration by parts identity for the central difference. $\Box$

    A number of FT brackets have closely similar LQG
analogs.  The following example is given without proof (FT bracket first; then analogous
SS LQG bracket).
\begin{align}
     \{\mathrm{K E}(z),\partial_z ' \mathrm{E}(z')/\mathrm{E}(z')\} &=  [\partial_z'\delta(z-z')]\mathrm{E}(z)/\mathrm{E}(z')
                     - \delta(z-z') \partial_z ' \mathrm{E}(z')/\mathrm{E}(z') \nn
                   & =  \partial_z'\delta(z-z') \quad \mbox{(FT)}; \nn
      \{\mathrm{K E}(q),\cd \mathrm{E}(m)/\mathrm{E}(m)\} &= \cd(m) \delta(q,m)\mathrm{E}(q)/\mathrm{E}(m)
                 - \delta(q,m) \cd \mathrm{E}(m)/\mathrm{E}(m)\nn
                 &= \cd(m) \delta(q,m)  \quad \mbox{(SV)}
 \label{KEwithcdEOverE}
\end{align}
K and E stand for any $\rK^I_j$ and its conjugate $\rE^j_I$.  The quantity
 \[
    \cd(m) \delta(q,m) := [\delta(q,m+1) - \delta(q,m-1)]/2,
 \]
is the difference of a Kronecker delta  $\delta(q,m)$.  This
difference is the discrete analog
of the derivative of a Dirac delta function.

\section{Constraints in the small sine limit}
\label{Gauss}

\subsection{The Gauss constraint}

       In QFT, the Gauss Identity
\begin{equation}
   0 = \partial_i \mathrm{E}^i_I + \epsilon_{IJK} \rA^J_m \rE^m_K
\label{QFTGauss}
\end{equation}
can be broken into two parts,
\begin{eqnarray}
    0 &= & \partial_i \rE^i_I + \epsilon_{IJK}\Gamma^J_m \mathrm{E}^m_K ; \nn
    0 &=& \epsilon_{IJK} \mathrm{K}^J_m \mathrm{E}^m_K.
\label{PartsofGaussIdentity}
\end{eqnarray}
The breakup is possible because the first line vanishes
by itself: the covariant divergence
of a density one triad vanishes, and  involves
no Christoffel symbols.

    In the plane wave
case only the U(1) corresponding to rotations around
the Z axis survives.  Line one above becomes
\begin{equation}
    0 =  \cd \mathrm{E}^z_Z + \epsilon_{ZAB} \, \Gamma^A_m \, \mathrm{E}^m_B.
\label{SSEzIdentity}
\end{equation}
The derivative has been replaced by a central difference.
\Eq{SSEzIdentity} may be
derived from the SS $\Gamma$, \eq{SSKGamma}.
Main steps in a direct proof: relabel M,I $\rta$ J,K on the
last line of \eq{SSKGamma}, and replace the $\Sigma$ by triads, using
\begin{eqnarray}
    \Sigma^{ijK} &:=& \mathrm{E}^i_I \mathrm{E}^j_J \epsilon^{IJK}/|e|  \nn
                &=& e^i_I e^j_J \mid e \mid \nn
                &=& \si(e) \, e^K_k \epsilon^{ijk}.
\label{SigmaEqualsTriads}
\end{eqnarray}
Then use the antisymmetry of the
Levi-Civita tensors to replace
\begin{eqnarray}
    \epsilon^{zmr} \cd e^J_r  \,e^{I}_m &\rta& (\epsilon^{zmr}/2)\,[ \, (\cd e^J_r)  \,e^{I}_m + (\cd e^I_m ) \,e^{J}_r \,]\nn
    &\cong& (\epsilon^{zmr}/2) \, \cd (\, e^J_r e^{I}_m \,)\nn
    &=& (1/2)\, \si(e)\, \cd \rE^z_Z \quad \Box
\label{GaussIdentity2}
\end{eqnarray}
The second line of \eq{PartsofGaussIdentity} must
be imposed as a constraint on the Hilbert space
of coherent states.

\subsection{The Lorentzian Hamiltonian}
\label{LorentzH}

    The Lorentzian Hamiltonian H equals minus the Euclidean
Hamiltonian, plus terms quadratic in the extrinsic curvature.
\begin{eqnarray}
    \mathrm{H} & = & \sum_n [-(1+ \gamma^2)/2\kappa]\, (\mathrm{K}_i^I \,\mathrm{K}_j^J \,\epsilon_{IJK} \,e^{ijK} \rN)(n) - \mathrm{H}_e.  \quad \mbox{(QFT)}
\label{ConstructH}
\end{eqnarray}

    The Hamiltonian of \eq{ConstructH}
contains three variables: $\mathrm{K}_i,\,  \hat{h}_i$, and
$\Sigma^{ijK}$. They are not independent, and one must decide
which variable to eliminate. From \eq{SSKGamma}, one can eliminate
either $\mathrm{K}_i$ or $\hat{h}_i$.  Either choice introduces
a new, and complicated field, the $\Gamma_i$.

    There is no way of avoiding the $\Gamma_i$.  However,
the unidirectional constraints will allow K to be replaced by
a function of the $\Sigma$.   Anticipating this, we eliminate
the $\hat{h}_i$.
\begin{eqnarray}
    \mathrm{H} +\mathrm{ST}& =&  \sum_n [1/\kappa]\{-\mathrm{K}^I_x \, \mathrm{K}^J_y \, \epsilon_{IJK} \, \epsilon^{KMN} \, \mathrm{E}^x_M \, \mathrm{E}^y_N \,N(n)/\mid e \mid   \nn
              & &  - \mathrm{K}^Z_z \,\mathrm{K}^A_a  \,\mathrm{E}^a_A \,\mathrm{E}^z_Z N(n) /\mid e \mid \nn
              & &+ \Gamma^I_x \, \Gamma^J_y \, \epsilon_{IJK} \,\mathrm{E}^x_M \, \mathrm{E}^y_N \, \epsilon^{MNK} \, N(n)/ \mid e \mid  \nn
               & & + \Gamma^Z_z \, \Gamma^A_a \, \mathrm{E}^a_A \,\mathrm{E}^z_Z \, N(n) /\mid e \mid  \nn
        & & -\Gamma^A_a \, \epsilon_{BA}\, \cd [\, N \, \mathrm{E}^z_Z \mathrm{E}^a_B(n)/\mid e \mid\,] \}.
\label{H1}
\end{eqnarray}

\subsection{Identities obeyed by the $\Gamma$}

    In \eq{H1}, terms linear in K have canceled out, because of
Gauss, second line of \eq{PartsofGaussIdentity}, plus an
identity obeyed by the $\Gamma$,
\begin{equation}
    0 = \cd  \,\Sigma^{mzI} + \epsilon_{IJK} \,\Gamma_i^J  \,\Sigma^{miK}.
\label{SSPartiale}
\end{equation}
This is the LQG analog of a FT identity: with $\cd$ replaced
by $\partial_z$, \eq{SSPartiale} states that the covariant
divergence of a function of triads must vanish.
The relation involves no Christoffel symbols because of
the antisymmetry in indices z, a.  The FT version of this
identity guarantees that the FT Hamiltonian contains no
linear in K terms; the LQG version functions similarly.

    \Eq{SSPartiale} can be used to \emph{define} the $\Gamma$, since it can be
solved for the $\Gamma$.  The $\Gamma$ obtained in this manner
are the  same as the $\Gamma$ obtained from Thiemann's procedure, \eq{SSKGamma}.
Main steps in the proof: multiply the above equation by the triad $e^M_m$. In the second, $\Gamma\Sigma$
term, the $\Sigma^{miK}e^M_m$ gives an \Etld.  In the first, $ \cd  \,\Sigma$
term, replace $e^M_m$ by a $\Sigma$, by inverting
the relation between $\Sigma$ and triad:
\begin{eqnarray*}
    \Sigma^{imM} &:=& \rE^i_I \rE^m_N \epsilon^{INM}/ \mid e \mid \nn
                 & =& e_r^M \epsilon^{rim} \, \si(e); \nn
    e^M_m &=& \si(e) \, \Sigma^{inM} \epsilon^{min}/2!.\; \Box
\end{eqnarray*}

    The following relations are also useful for
simplifying the Hamiltonian.
\begin{eqnarray}
    \Gamma_z^A &=& \Gamma_a^Z = 0;
\label{GammaBlockDiag} \\
    \Gamma_a^A \rE_A^a  &=& 0; \quad \mbox{a, A transverse}.
\label{TransverseTraceVanishes}
\end{eqnarray}
I.~e. $\Gamma$, like \Etld, is block diagonal,
with $\Gamma^Z_z$ in the 1 x 1 block;
and the transverse trace vanishes.  Both these
results follow from \eq{SSKGamma} by taking M = Z, A
and I = Z, A in turn.

    The $\Gamma$ obey various relationships in FT,
and \emph{most of these remain valid in SS LQG,}
despite replacement of derivatives by differences.  This is
so, because the relationships are proved using only algebra.
The  proofs  do not involve calculus or the properties of the derivative.

\subsection{The vector constraint  $H_z$}

    Since the wavefunctional to be constructed
is not based on closed loops, the diffeomorphism
constraint is not satisfied automatically.  It must be
treated as an additional constraint.

    The classical constraint is
\begin{equation}
    \mathrm{N}^z \mathrm{H}_z = (1/\kappa \gamma)\int d^3x \mathrm{N}^z \,\mathrm{F}_{za}^A \,\mathrm{E}^a_A. \quad \mbox{(QFT)}
\end{equation}

    Make the same assumptions as for the Hamiltonian:
the non-local version includes both nearest neighbors with
equal weight; small sine and slow variation approximations apply.
The outcome is as for the Hamiltonian: the  $\hat{h}_z(n_z, n_z\pm 1)$ are replaced
by an average; the $\hat{h}_a(n_z \pm 1)$ are replaced by $\hat{h}_a(n_z)$;
derivatives are  replaced by central differences.
\begin{eqnarray}
     \kappa \gamma\mathrm{N}^z \mathrm{H}_z
            &=& \sum_n \mathrm{N}^z (Tr/2)\{(-2 i) \,\cd \hat{h}_a(n) \,\mathrm{E}^a(n) \nn
            &&\quad +2 i [\hat{h}_z,\hat{h}_a(n)] \,\mathrm{E}^a(n)\}.\quad \mbox{(SS, SV)}
\label{Hz2}
\end{eqnarray}

    As in the FT case, one must add in a term proportional to the Gauss
constraint to make $\mathrm{H}_z$ into the generator of z diffeomorphisms.
The second line of \eq{Hz2} equals
\begin{equation}
     (-2 i)^2 \, \hat{h}^Z_z \, \epsilon_{AB} \,\hat{h}^A_a(n) \,\mathrm{E}^a_B(n).
\label{HzContainsGauss}
\end{equation}
ONe can replace $(-2i) \, \hat{h} \,\rta \,\gamma \mathrm{K} + \Gamma$.
From Gauss, \eq{PartsofGaussIdentity}, the term involving
K vanishes, and the term involving $\Gamma$
equals one half of the
Gauss constraint, \eq{PartsofGaussIdentity}.
\Eq{Hz2} becomes
\begin{eqnarray}
    \kappa \gamma\mathrm{N}^z \,\mathrm{H}_z &=& \sum_n \mathrm{N}^z \{(-2 i) \,\cd\hat{h}^A_a(n) \,\mathrm{E}^a_A(n) \nn
        &&\quad -(-2i) \,\hat{h}_z(n) \,\cd \mathrm{E}^z_Z\}.
\label{Hz}
\end{eqnarray}
The next few sections simplify the Hamiltonian
by choosing gauges and imposing constraints.

\section{Single polarization constraints}
\label{SinglePol}

    The single polarization constraints are
\begin{equation}
    \mathrm{E}^x_Y = \mathrm{E}^y_X = 0.
\label{SinglePolarization}
\end{equation}
These constraints must obey the consistency conditions
\[
    \{ \,\mathrm{H + ST},\rE^x_Y \,\} = \{ \,\mathrm{H + ST},\rE^y_X \,\} = 0,
\]
which of course just require the vanishing of the
conjugate coordinates.
\begin{equation}
    \mathrm{K}^x_Y = \mathrm{K}^y_X = 0.
\label{OffDiagKVanish}
\end{equation}

    For brevity these conditions will be denoted
simply as "single polarization" constraints; but they
not only specialize to a single
polarization; they also fix the U(1) gauge.
If one wishes to specialize to single
polarization, without fixing the U(1)
gauge, one may impose \E{x}{I} \E{y}{I} = 0.

    The triad and \Etld matrices are now
diagonal.  Additionally, from \eq{SigmaEqualsTriads},
all three indices of $\Sigma^{ijK}$ must be
unequal. For example,
\[
    \Sigma^{mzZ} = \Sigma^{xyX} = 0, \quad \mbox{(single pol.)}
\]
while $\Sigma^{xyZ}$ is finite.
This follows  from \eq{SigmaEqualsTriads} and the
diagonal nature of the triads.
Also, from \eq{SSKGamma}, the only
surviving $\Gamma$ are the two off-diagonal $\Gamma^X_y, \Gamma^Y_x$.
\begin{equation}
    \Gamma^X_x = \Gamma^Y_y = \Gamma^Z_z = 0.\quad \mbox{(single pol.)}
\label{GammaOffDiagonal}
\end{equation}

\section{ The diffeomorphism constraint}
\label{DiffGauge}

    In both FT and LQG the usual gauge choice which fixes
the Lorentz boosts, reducing the full Lorentz
group to SU(2), is

\begin{equation}
    e^t_{X,Y,Z} = 0 = e^{x,y,z}_T .
\end{equation}
This gauge still allows
transformations

\begin{equation}
    t' = t'(t); \qquad z' = z'(z,t)
\end{equation}

    The transverse triads vary with this change in the
z coordinate, despite their lack of an explicit z index,
because e, the volume factor, contains an implicit
z subscript.   Conversely, the longitudinal triad (has
an explicit z index but) does not change with change in
z coordinate.

\begin{eqnarray}
    \mathrm{E}^a_A &\propto& |e| = \si(e)\, e^Z_z \,^{(2)}e; \nn
    \mathrm{E}^z_Z &= & |e| \,e^z_Z = \si(e)\, (^{(2)}e);\nn
\end{eqnarray}
$^{(2)}e$ is the determinant of the 2x2 transverse
triad matrix, an invariant.  Therefore $\mathrm{E}^z_Z$ is a scalar,
while the $\mathrm{E}^a_A$ are rank one covariant tensors.
In FT, therefore, a gauge
fixing constraint must involve at least some transverse
triads.
     We use a gauge fixing function constructed from the
two simplest triad functions which are U(1) scalars, \Etwo and
\E{z}{Z}.

\begin{eqnarray}
    0 &=& \ln \,[\Etwo/(\rC\mathrm{E}^z_Z)^{p+1/2}] := \mathrm{D}_1; \nn
      0 &=& 2 \, \mathrm{K}_z \,\mathrm{E}^z + \mathrm{K}_a \,\mathrm{E}^a/2 -p \,\mathrm{K}_a \,\mathrm{E}^a := \mathrm{D}_2.
\label{RightGaugeChoice}
\end{eqnarray}
C is a constant.  \Etwo is the determinant of the 2x2 transverse
cotriad matrix. The single polarization constraints imply
\[
    \Etwo = \rE^x_X \rE^y_Y.
\]
 \Eq{RightGaugeChoice} is a family of gauge
choices, depending on a parameter p.  $\mathrm{D}_1$ depends on p +1/2,
rather than p, because at a later point the value
p = 0 will prove to be special.  The
second line is the consistency condition,
the result of demanding \{H, $\mathrm{D}_1$\} = 0.

    \Eq{RightGaugeChoice} is by no means the only way
of fixing the diffeomorphism gauge.  However, it is the
simplest.  More complex choices found in the literature
appear to require advance knowledge of the form of the
solution.

    The first line proposes a gauge choice involving a logarithm,
rather than a simpler choice

\begin{equation}
    \Etwo - (\mathrm{C} \,\mathrm{E}^z_Z)^{p+1/2} = 0.
\label{D3}
\end{equation}
To see the reason for this, recall fixing the
diffeomorphism gauge is equivalent
to first transforming to new canonical coordinates ($\pi$,q), then
discarding one pair of ($\pi$,q)'s.  Write
\[
    \mathrm{E}^i_I \,(\rmd \mathrm{K}^I_i/\rmd t) = -\rmd \mathrm{E}^i_I/\rmd t \,\mathrm{K}^I_i + \:\mbox{total
derivative},
\]
then expand:
\begin{align}
     -\mathrm{K}^i_I \rmd\mathrm{E}^i_I/\rmd t &= -\mathrm{K}_i \mathrm{E}^i \rmd(\ln \mathrm{E}^i)/\rmd t \nn
                        &= -\mathrm{K}_z \,\mathrm{E}^z \, \rmd(\ln \mathrm{E}^z)/\rmd t - (1/2)\mathrm{K}_a \mathrm{E}^a \rmd(\ln \Etwo)/\rmd t\nn
                        &\quad  -(1/2)(\mathrm{K}_y \,\mathrm{E}^y - \mathrm{K}_x \,\mathrm{E}^x) \rmd[\ln (\mathrm{E}^y/\mathrm{E}^x)]/\rmd t   \nn
                        &= +\{(2 \,\mathrm{K}_z \,\mathrm{E}^z + \mathrm{K}_a \,\mathrm{E}^a/2 -p \, \mathrm{K}_a \,\mathrm{E}^a) \nn
                        &\quad \times \rmd[\ln \Etwo -(\ln \mathrm{E}^z)/2 -p \ln \mathrm{E}^z]/\rmd t -(p\rta-p)\}(1/4p)\nn
                        &\quad -(1/2)(\mathrm{K}_y \mathrm{E}^y - \mathrm{K}_x \mathrm{E}^x) \, \rmd[ \, \ln (\mathrm{E}^y/\mathrm{E}^x) \,]/\rmd t.
\label{CanonicalTransformation}
\end{align}
The fourth and fifth lines are $\mathrm{D}_2$ times the derivative
of  $\mathrm{D}_1$.  One can drop this ($\pi$,q)
pair completely from the theory,
without altering the canonical brackets of
the other ($\pi$,q) pairs. The constant
C of \eq{D3} arises as a constant of integration.  Note the special
case p = 0 has a singularity.

    In practice the gauge choice \eq{RightGaugeChoice} does not introduce
logarithms into the constraints; only
\eq{D3} and its first difference are needed
when simplifying the Hamiltonian.

    The popular choice For C and p, in the
classical literature, is C = sgn(e), p = 1/2,
which implies $\mathrm{g}_{zz}$ = 1.
C and p will be determined in
the succeeding paper.

    The case p = 0 clearly requires a
special discussion.  Because the classical literature
favors the gauge choice p = 1/2 , presumably
the p = 0 case will not be needed.  This paper does not discuss it.

\section{The switch from N to \underline{N}}
\label{Nt}

\subsection{Boundary conditions at infinity}

    In Newtonian static planar gravity
the gravitational potential at infinity
does not die off as some power of z, but rather
grows linearly. In general relativity, the Newtonian
result does not rule out flat space at infinity, because an observer
in a free-fall elevator would detect the \emph{same} force
at top and bottom of the elevator.   Attempts to
generalize the Newtonian static result to full general
relativity have not been successful \cite{NoStatic}.
Reasonable restrictions on the stress-energy of the planar matter source
presumably lead to instability.

    If the source is a time-varying wave packet,
Presumably it is safe to assume flatness at infinity,
because of causality: the packet has not yet reached
infinity.  However, for now there is no loss of
generality if one makes the more conservative
assumption, conformal flatness at infinity.
The (z,t) portion of the metric at infinity is
assumed to take the conformal form

\begin{eqnarray}
     [-\mathrm{N}^2 + (\mathrm{N}^z)^2 g_{zz}] dt^2 &+ 2 \mathrm{N}^z g_{zz}dz dt + g_{zz} dz^2 \nn
     & \,\rta g_{zz} \,(-dt^2 + dz^2),
\label{ConformalForm}
\end{eqnarray}
where N and $\mathrm{N}^z$ are the ADM
lapse and shift.  This requires the boundary conditions

\begin{eqnarray}
    \mathrm{N}^z &\rta& \, 0;\nn
    \Nt^2 &:=& \mathrm{N}^2/g_{zz} \rta \, 1.
\label{BoundaryConditions}
\end{eqnarray}
\Nt, rather than N, goes to $\pm 1$.
The underlining is needed because \Nt is density weight -1.
Tildes are not always used to indicate the density weight of the
triads.  They are familiar to most readers,
and it is understood the triads are weight 1.
\Nt, however, is an unfamiliar quantity, and its
density weight will play a role in section \ref{Unidirectional},
when the unidirectional constraints are imposed.

    For plane waves, the boundary conditions require a
shift from  N to \Nt.  In cotriad notation,
\begin{equation}
    \Nt(n) := (\rN \,\mathrm{E}^z_Z/ \mid e \mid)(n)
\label{defNt}
\end{equation}  The lapse N is a scalar under
spatial diffeomorphisms.
Therefore from \eq{defNt} \Nt is a rank one contravariant
tensor.  N has no factors
of $\Delta x^i$, but \Nt has a factor 1/$\Delta z$.

    This shift in lapse generates a shift in the Hamiltonian.
\begin{eqnarray*}
    \mathrm{N} \,&\rH &:= \Nt \,\Ht; \\
    &\Ht &= \mathrm{H} \,\mid e \mid/\mathrm{E}^z_Z.
\end{eqnarray*}
When the $\Sigma$ are expressed in terms of the \Etld,
and N is replaced by \Nt, the $1/\mid e\mid$ singularities
disappear, but some terms acquire a 1/\E{Z}{z} singularity.
\begin{eqnarray}
    \mathrm{N \,H}_e \rta \Nt \,\Ht_e &=& \sum_n \Nt \,\{\mathrm{F}^Z_{xy} \epsilon^{ZAB}\mathrm{E}^x_A \mathrm{E}^y_B/\mathrm{E}^z_Z \nn
                                    &&\quad  + \mathrm{F}^A_{za}\epsilon^{AZB} \mathrm{E}^a_B\}/\kappa + \mathrm{ST}.
\label{HeWithNt}
\end{eqnarray}
The Lorentzian Hamiltonian, \eq{H1}, becomes
\begin{align}
    \mathrm{NH +ST} &=  \sum_n (1/\kappa)\{- (\mathrm{K}^A_a \,\mathrm{E}^a_A)^2 \,\Nt(n)/(4 \,\mathrm{E}^z_Z )\nn
                    \quad &   + (\mathrm{K}^Y_y \,\mathrm{E}^y_Y - \mathrm{K}^X_x \,\mathrm{E}^x_X)^2 \Nt(n)/(4 \,\mathrm{E}^z_Z)  \nn
              \quad & -\mathrm{K}^Z_z(n) \,\mathrm{E}^z_Z  \,\mathrm{K}^A_a(n) \,\mathrm{E}^a_A \,\Nt(n)/ \mathrm{E}^z_Z\nn
              \quad & +(\Gamma^Y_x \,\mathrm{E}^x_X +\Gamma^X_y \,\mathrm{E}^y_Y)^2(n) \,\Nt(n) /( 4\, \mathrm{E}^z_Z) \nn
               \quad & -(\Gamma^Y_x \,\mathrm{E}^x_X -\Gamma^X_y \,\mathrm{E}^y_Y)^2(n) \,\Nt(n) /( 4 \,\mathrm{E}^z_Z) \nn
       \quad & -\Gamma^A_a(n) \,\epsilon_{BA} \,\cd (\mathrm{\Nt \mathrm{E}^a_B}) \}.
\label{H2}
\end{align}
This equation incorporates the single polarization constraints (K and E entirely diagonal,
$\Gamma$ entirely off-diagonal), but not the diffeomorphism constraints.
However, the equation anticipates those constraints and pairs each K with an E.
(The reshuffled E's are still to the right of their conjugate K's.)

    \Eq{H2} also switches to combinations of $\Gamma$ and E which are
relatively simple to express in terms of \Etld.
\begin{eqnarray}
    \Gamma^Y_x \,\mathrm{E}^x_X + \Gamma^X_y \,\mathrm{E}^y_Y &=& [\cd \mathrm{E}^y_Y/\mathrm{E}^y_Y - \cd \mathrm{E}^x_X/\mathrm{E}^x_X] \,e^X_x e^Y_y \si\nn
    &=& [\cd \mathrm{E}^y_Y/\mathrm{E}^y_Y - \cd \mathrm{E}^x_X/\mathrm{E}^x_X] \,\mathrm{E}^z_Z;\nn
     \Gamma^Y_x \mathrm{E}^x_X - \Gamma^X_y \mathrm{E}^y_Y &=& \cd \mathrm{E}^z_Z.
\label{GammaEqualsTriads}
\end{eqnarray}
Proof of \eq{GammaEqualsTriads}:    from \eqs{SSKGamma}{SigmaEqualsTriads},
\[
    \Gamma^X_y \,\mathrm{E}^y_Y = -\si(e) \,\cd e_n^X \,\epsilon^{zmn} \,e^Y_m \quad
            \mbox{(single pol.)},
\]
plus an additional formula with X $\leftrightarrow$ Y.

    If the equation $e^Y_y \,$ \E{y}{Y} = $\mid e \mid$ is differenced, then
divided by  $e^Y_y$ \E{y}{Y}, one gets
\[
    \cd e^Y_y /e^Y_y + \cd \mathrm{E}^y_Y/\mathrm{E}^y_Y = \cd \mid e \mid/\mid e \mid,
\]
plus a similar equation for x $\rta$ y.  These equations imply the first
line of \eq{GammaEqualsTriads}.  The last line is one half of Gauss,
\eq{SSEzIdentity}.

    In quantum geometrodynamics for the plane wave case, the shift from N to \Nt
modifies the constraint algebra, making it anomaly-free.
Usually, a commutator of constraints produces a metric component
to the right of a constraints, causing an anomaly.  When N is
replaced by \Nt, the dangerous metric component  is
absorbed into \Nt and disappears from the commutator. \footnote{ The author would appreciate
help from readers in locating the original source  of this
result: no anomalies in the planar case.}

\section{The unidirectional constraints}
\label{Unidirectional}

\subsection{Are Dirac brackets necessary?}

    Since unidirectional constraints typically are second class,
it is necessary to replace Poisson by Dirac brackets.
Dirac brackets often are not pretty.  Are there ways of
avoiding the introduction of Dirac brackets?

    At the classical level, one of the unidirectional
constraints is satisfied, if the remaining two unidirectional
constraints plus scalar and vector constraints are satisfied \cite{HintMaj}.
Also, one can write one of the
unidirectional constraints as a linear combination of the
diffeomorphism constraints.  This suggests only one
unidirectional constraint survives, so it must be first
class (commute with itself).

    Eliminating a unidirectional constraint by writing
it as a combination of diffeomorphism constraints does simplify
the calculations, but does not eliminate completely the need
for Dirac brackets.  Even if only one unidirectional constraint
survives, it does not commute with itself.  The generic
unidirectional constraint has the form $ \mathrm{U} = \pi + \cd q = 0$
(time derivative plus z derivative vanishes).  The generic commutator is
\begin{eqnarray}
    [\mathrm{U}(z_1),\mathrm{U}(z_2)]&=&[\pi + \partial_{z1} q (z_1), \,\pi + \partial_{z_2} q (z_2)] \nn
    &=& (-i\hbar) \, [+\partial_{z2}\delta(z_1-z_2) - \partial_{z1}\delta(z_1-z_2)]  \nn
    &=& -2 \,i \, \hbar \,\partial_{z2}\delta(z_1-z_2).
\label{UnidirectionalCommutator}
\end{eqnarray}
The calculation is done in FT for the convenience of the
reader, but the result in LQG is similar: replace
\begin{eqnarray}
    (z_1,z_2) \:&\rta &\: (n_1,n_2);\nn
    \partial_{z2} \,\delta(z_1-z_2) &\rta &\:[ \,\delta(n_1,n_2+1) - \delta(n_1, n_2-1) \,]/2 \nn
                       & := &\cd(n_2) \,\delta(n_1,n_2)
\label{defDifferenceofDelta}
\end{eqnarray}
The derivative of a Dirac delta is replaced by
the difference of a Kronecker delta.
The commutator of a unidirectional constraint with
itself does not vanish.  The discrete version of the
commutator suggests a reason for this behavior: the $\cd$q
term is non-local.

    The non-linearity of general relativity
has nothing to do with the foregoing result. Even in the
simplest, linear
field theory (real scalar free field) a unidirectional
constraint will be second class.   It is not possible to
avoid Dirac brackets.

\subsection{The unidirectional operator}

    In conventional wave theory a solution is
unidirectional if all fields depend only on
z - ct.  In general relativity those coordinates
are arbitrary, and one must use local
free-fall coordinates Z - cT instead.

    In terms of derivatives
\begin{eqnarray*}
    \sqrt{2}\partial_U = \partial_Z-\partial_T ;\\
    \sqrt{2}\partial_V = \partial_Z+\partial_T.
\end{eqnarray*}
The constraint (no V dependence) is  $(\partial_Z+\partial_T) = 0$.

    This constraint can be rewritten in terms of (z,t) derivatives.
\begin{eqnarray}
    0 &=& (\partial_Z+\partial_T) \, f(Z-cT) \nn
    &=& (e^z_Z \, \partial_z + e^t_Z \, \partial_t + e^z_T \, \partial_z + e^t_T \, \partial_t)f \nn
    &=& ( e^z_Z \, \partial_z + 0 + (-\mathrm{N}^z/\mathrm{N}) \, \partial_z + (1/\mathrm{N}) \,\partial_t)f.
\label{UnidirectionalConstraint1}
\end{eqnarray}
This equation invokes the usual gauge
which fixes the Lorentz boosts and reduces the full
Lorentz group to SU(2): $e^t_{X,Y,Z} = 0$.

    One can replace
the  derivatives $\partial_z$ and $\partial_t$
in \eq{UnidirectionalConstraint1}
by Poisson brackets with  $\mathrm{H}_z$
and $\Nt \Ht(z)+\mathrm{ST}+\mathrm{N}^z\mathrm{H}_z$ respectively.

\begin{eqnarray}
    0 &=& [ \, e^z_Z -\mathrm{N}^z/\mathrm{N}(z) \,]\{f,\mathrm{H}_z(z)\} \nn
        &&\quad + (1/\mathrm{N})\{\,f, (\Nt \,\Ht(z)+\mathrm{ST}) + \mathrm{N}^z \,\mathrm{H}_z(z)\} \nn
    & \propto &  \, (\mathrm{N} \mathrm{E}^z_Z/\mid e \mid) \, \{f, \mathrm{H}_z(z)\} + \{ f, ( \Nt \,\Ht(z)+\mathrm{ST}) \}  \nn
    &=&  \{f,\Nt  \,\mathrm{H}_z(z) +  (\Nt \,\Ht(z)+\mathrm{ST})\}.
\label{UnidirectionalConstraint2}
\end{eqnarray}
The third line is multiplied by N.  this anticipates
a later result: given the unidirectional
constraints and the diffeomorphism gauge choice, N
cannot vanish.

    Strictly speaking $\mathrm{H}_z$ is not the z derivative
operator unless its Lagrange multiplier \Nt is a
constant.  If \Nt is not a constant, $\{f, \mathrm{H}_z \}$ does take
the z derivative of f; but it also generates gauge transformations
proportional to $\partial_z \,\Nt$.  In the
present case this  is  not a problem;
once unidirectional and diffeomorphism constraints are imposed,
\Nt will turn out to be a constant.

    The constraint \eq{UnidirectionalConstraint2} may be written out
explicitly, using \eqs{Hz}{H3} for $\mathrm{H}_z$ and \Ht.

\begin{align}
   \Nt  \mathrm{H}_z(z) +  \Nt \,\Ht(z)+\mathrm{ST}&= (1/\kappa )\sum_n \Nt\{(-2 i) \,\cd\hat{h}^A_a(n) \,\mathrm{E}^a_A(n) \nn
        & -(-2i) \,\hat{h}_z(n) \,\cd \mathrm{E}^z_Z\} \nn
        & + \sum_n (1/\kappa)\{ -\mathrm{K}^X_x \,\mathrm{K}^Y_y  \,\mathrm{E}^x_X \,\mathrm{E}^y_Y \,\Nt(n)/\mathrm{E}^z_Z  \nn
              &   - \mathrm{K}^Z_z(n) \,\mathrm{K}^A_a(n) \,\mathrm{E}^a_A(n) \,\Nt(n) \nn
              &  +(\Gamma^Y_x \,\Gamma^X_y)(n) \,\Nt(n) \,\Etwo(n)/\mathrm{E}^z_Z \nn
        &  -\Gamma^A_a(n) \,e^{zaA}(n) \,\cd \mathrm{N} \}.
\label{ExplicitUniConstraint0}
\end{align}

\subsection{The unidirectional constraints}

    In a unidirectional theory, \eq{ExplicitUniConstraint0}
commutes with every dynamical variable.  Therefore one can
construct unidirectional constraints by
commuting \eq{ExplicitUniConstraint0} with
any set of independent functions $f_i$, then setting
the commutators equal to zero.  Choose the $f_i$
to be three independent
functions of the three triads:
\E{z}{Z}, \Etwo, and $\ln$[\E{y}{Y}/\E{x}{X}].
Commutation of these three yields
the constraints
\begin{eqnarray}
    0 &=& \{\mathrm{K}^A_a \,\mathrm{E}^a_A + \cd \mathrm{E}^z_Z \}/\sqrt{\mathrm{E}^z_Z}:= \mathrm{U}_1;\nn
    0 &=&  \{\mathrm{K}^A_a \,\mathrm{E}^a_A + 2 \,\mathrm{K}^Z_z \,\mathrm{E}^z_Z  \nn
    &&\quad + \mathrm{E}^z_Z \,\cd \Etwo/\Etwo + 2 \,\mathrm{E}^z_Z \,\cd \Nt/\Nt\}/\sqrt{\mathrm{E}^z_Z} := \mathrm{U}_2;\nn
    0 &=& \{\mathrm{K}^Y_y \,\mathrm{E}^y_Y - \mathrm{K}^X_x \,\mathrm{E}^x_X \nn
    &&\quad - \mathrm{E}^z_Z \,[\cd \mathrm{E}^y_Y /\mathrm{E}^y_Y - \cd \mathrm{E}^x_X/ \mathrm{E}^x_X]\}/\sqrt{\mathrm{E}^z_Z} := \mathrm{U}_3.
\label{NaiveUniConstraints}
\end{eqnarray}
The appearance of $\cd \Nt$ in the second constraint may be a
bit surprising.   This comes from a bracket
\begin{align*}
   \{\Nt (-2i)\,\cd \hat{h}\,\mathrm{E},\Etwo\} &= \{\sum_n \Nt\{(-2 i) \,\cd\hat{h}^A_a(n) \,\mathrm{E}^a_A(n), \Etwo(m)\} \nn
            &= - \cd (\Nt \,\mathrm{E}^y_Y) \,\mathrm{E}^x_X + (x\leftrightarrow y, X \leftrightarrow Y),
\end{align*}
followed by multiplication by $\sqrt{\mathrm{E}^z_Z}/\Nt \,\Etwo$.
In effect, the $\cd$ has been integrated by  parts off the
holonomy and onto the \Nt and \Etld.  Similarly, an integration
by parts produces the $\cd \Etld/\Etld$ terms in the third
constraint; in that case a $\cd \Nt $ term cancels out.

    The usual lapse N
is a scalar, but the new lapse \Nt is a
contravariant tensor.  Both differences,
$\cd$\Etwo/\Etwo and  2 $\cd$\Nt/\Nt in
\eq{NaiveUniConstraints}, therefore have inhomogeneous terms in
their diffeomorphism transformation laws.  However, the
inhomogeneous terms cancel out in the sum.  The combination
which occurs in the unidirectional constraint,
\[
    \cd \Etwo/\Etwo + 2 \,\cd \Nt/\Nt,
\]
transforms like a tensor.

    The factors of 1/$\sqrt{\mathrm{E}^z_Z}$ have been added
to split up the 1/\E{z}{Z} singularity into two parts.  (The
Hamiltonian is  of the form $\sum \mathrm{U}_i \mathrm{U}_j$,
with  $\sqrt{1/\rE^z_Z}$ absorbed into each  $\mathrm{U}_i$.)

    These constraints have the right form.  The K dependent
terms represent time derivatives; the $\cd$ terms the
corresponding space derivatives.

\subsection{Eliminating one unidirectional constraint}
\label{ElinimateConstraint}

    Because of the diffeomorphism gauge fixing, the unidirectional
constraints $\mathrm{U}_1, \mathrm{U}_2$ are not independent.
For p $\neq$ 0,
one can solve \eq{RightGaugeChoice} or
\eq{D3} to eliminate a ($\pi$,q) pair.  The surviving ($\pi$,q)
pair is
\begin{eqnarray}
    \Pi &:=& (2 \,\mathrm{K}_z \,\mathrm{E}^z + \mathrm{K}_a \,\mathrm{E}^a/2 + p \,\mathrm{K}_a \,\mathrm{E}^a)/4p;\nn
    Q &:=& \ln \Etwo  + (p-1/2) \,\ln (\mathrm{C} \,\mathrm{E}^z_Z).
\label{defPiQ}
\end{eqnarray}
This is the pair indicated schematically by (p $\rta$ -p)
in \eq{CanonicalTransformation}.  The
factor 1/4p of that equation is absorbed into the $\Pi$
(an arbitrary choice); this gives bracket \{$\,\Pi$,\,Q\} the correct norm.

    Several quantities simplify when the diffeomorphism constraints
$\rD_i$ are dropped from the Hamiltonian.  In particular,
\begin{eqnarray}
    \mathrm{K}_a^A \,\mathrm{E}^a_A &=& (4p \,\Pi - \rD_2)/2p \,\rta \,2 \,\Pi;\nn
    2 \mathrm{K}_z^Z \,\mathrm{E}^z_Z + \mathrm{K}^a_A \,\mathrm{E}^a_A/2 &=& (4p \,\Pi + \mathrm{D}_2)/2 \,\rta \,2p \,\Pi;\nn
    2 \mathrm{K}_z^Z \mathrm{E}^z_Z &\rta& \:(2p-1)\Pi.
\label{KEEqualsPi}
\end{eqnarray}
Similarly for the triads,
\begin{eqnarray}
    \ln \Etwo &=& Q (p + 1/2)/2p + \rD_1(p-1/2)/2p \: \rta \: Q (p + 1/2)/2p ;\nn
    \ln (\mathrm{C} \,\mathrm{E}^z_Z) &=& (Q - \rD_1)/2p \: \rta \: Q/2p; \nn
    \ln (\Etwo/\mathrm{C} \,\mathrm{E}^z_Z) \: &\rta& \: Q (p-1/2)/2p.
\label{LnEEqualsQ}
\end{eqnarray}

    When \eqs{KEEqualsPi}{LnEEqualsQ} are inserted into the
unidirectional constraints, the two constraints
$\mathrm{U}_1,\mathrm{U}_2$ in \eq{NaiveUniConstraints} collapse to the
same constraint, except $\mathrm{U}_2$ has an extra term
proportional to $(\cd \Nt)/\Nt$.  Therefore
that expression must vanish.  In place of \eq{NaiveUniConstraints}
one gets
\begin{eqnarray}
    0 &=& (\cd \Nt)/\Nt; \nn
    0 &=& \{\mathrm{K}^A_a \,\mathrm{E}^a_A + \cd\mathrm{E}^z_Z  \}/\sqrt{\mathrm{E}^z_Z}= \{2 \,\Pi +  \cd \mathrm{E}^z_Z\}/\sqrt{\mathrm{E}^z_Z} = \mathrm{U}_1; \nn
    0 &=& \{\mathrm{K}^Y_y \,\mathrm{E}^y_Y - \mathrm{K}^X_x \,\mathrm{E}^x_X \nn
        && - [ \,\cd \mathrm{E}^y_Y /\mathrm{E}^y_Y - \cd \mathrm{E}^x_X/ \mathrm{E}^x_X \,] \,\mathrm{E}^z_Z\}/\sqrt{\mathrm{E}^z_Z}  = \mathrm{U}_3.
\label{UniConstraints}
\end{eqnarray}
Because of the conformal boundary conditions,
\Nt must equal unity.

\subsection{Dirac brackets}

    The two surviving unidirectional constraints  are still second
class. The Dirac bracket
matrix is (rows
and columns in order $\mathrm{U}_1, \mathrm{U}_3$)
\begin{equation}
\{\mathrm{U}_i(m),\mathrm{U}_j(n)\} = \left\{\begin{array}{cc}
                -(2/p) \,\mathcal{A} & \mathcal{C}\\
                -\mathcal{C} &-4 \, \mathcal{A} \\
                \end{array}
                \right\}
\label{BracketMatrix}
\end{equation}
\begin{eqnarray*}
\mathcal{A} &=& \cd(n) \,\delta(m,n) :=  (\delta(m,n+1) - \delta(m,n-1))/2 ;\\
\mathcal{C} &=& (1/2p) \, [ \,\cd \mathrm{E}^y_Y / \mathrm{E}^y_Y - \cd \mathrm{E}^x_X / \mathrm{E}^x_X \,] \,\delta(m,n).
\end{eqnarray*}
The inverse bracket matrix is
\begin{equation}
 \{\mathrm{U}_j(n),\mathrm{U}_k(r)\}^{-1}= \left\{\begin{array}{cc}
                            \mathcal{K}^{-1} & \mathcal{K}^{-1}\mathcal{C}\mathcal{A}^{-1}/4 \\
-\mathcal{A}^{-1}\mathcal{C}\mathcal{K}^{-1}/4, &-\mathcal{A}^{-1}/4 -\mathcal{A}^{-1}\mathcal{C}\mathcal{K}^{-1}\mathcal{C}\mathcal{A}^{-1}/16\\
                                            \end{array}\right\};
\label{InverseBracketMatrix}
\end{equation}
\begin{eqnarray*}
    \mathcal{K} &=& -(2/p) \, \mathcal{A} - \mathcal{C}\mathcal{A}^{-1}\mathcal{C}/4; \\
    \mathcal{A}^{-1}(n,r) &=& -\Theta(n-r).
\end{eqnarray*}

    The theta function is a discrete analog of the usual step function.
\begin{equation}
    \Theta(n-r)=\left \{\begin{array}{cl}
          0 & \quad \mbox{for} \quad n-r \mbox{ even, including}\, 0 \\
          +1&  \quad  \mbox{for}\quad n-r \mbox{ odd} > 0,   \\
         -1 &  \quad \mbox{for}\quad n-r   \mbox{ odd} < 0.
         \end{array}
         \right.
\label{defTheta}
\end{equation}
\begin{eqnarray}
    \cd(n) \,\Theta(n-r) &:=& \,[\Theta(n+1 -r) - \Theta(n-1-r)]/2 \nn
                        &=& \,\delta(n,r).
\label{cdTheta}
\end{eqnarray}
The last line is reminiscent of \eq{defDifferenceofDelta},
where the derivative of a Dirac delta is shown to have
a discrete analog, the difference of  a
Kronecker delta.  Similarly here, the continuous
formula
\[
    \partial_2 \Theta(z_1 - z_2) = \delta(z_1-z_2)
\]
has a discrete analog, \eq{cdTheta}.

    The solution for $\Theta$, \eq{defTheta}, is determined only
up to a solution to the homogeneous version of \eq{cdTheta}.
\begin{equation}
 \Theta(n-r)_h=\left \{\begin{array}{cl}
          a & \quad \mbox{for} \quad n-r \mbox{ even, including}\, 0 \\
            b &  \quad  \mbox{for}\quad n-r \mbox{ odd},
         \end{array}
         \right.
\label{defThetah}
\end{equation}
a and b constants.  In a scalar free field theory, if one
drops all the k $<$ 0 Fourier components from the field $\phi$,
the resulting commutator $[\phi(x) , \phi(y)] $ is a step function
which changes sign at x = y.  I have chosen $\Theta_h$ so that
the function $\Theta(n-r)$ also changes sign at n-r = 0.

    \E{z}{Z} and $\mathrm{K}^Z_z$ have disappeared
at this point, replaced by Q and $\Pi$.  However,
it is perhaps better to retain the more
compact and familiar \E{z}{Z}, rather than Q.  The
\{\E{z}{Z} $\Pi$\} bracket is (recall $(\E{z}{Z} \propto \exp[Q/2p])$)
\begin{equation}
    \{\E{z}{Z},\Pi\} = (1/2p)\E{z}{Z}.
\label{ExPiBracket}
\end{equation}
Again, the case p = 0 requires
separate discussion.

   The matrix of constraints contains a field-dependent quantity
\[
  \mathcal{C} = (1/2p) \,[ \,\cd \mathrm{E}^y_Y/ \mathrm{E}^y_Y - \cd \mathrm{E}^x_X/ \mathrm{E}^x_X \,] \,\delta(1,2).
\]
This is very unusual.  In most field theories
brackets between unidirectional  constraints are field-independent. In weak field
limit,  $\mathcal{C}$ disappears from the Dirac brackets
because \E{z}{Z}  $\rta$ 1, and the off-diagonal bracket
\[
    \{\,\mathrm{U}_1,\mathrm{U}_3\,\} \sim \{ \,\Pi, \sqrt{E^z} \,\}
\]
vanishes.  The presence of $\mathcal{C}$ is therefore a
consequence of the nonlinearity of the theory, as represented
by the area factors \E{z}{Z}.

    The field dependence prohibits an exact solution
for $\mathcal{K}^{-1}$.  However, an integral equation
for $\mathcal{K}^{-1}$ has a power series solution.
\begin{eqnarray}
    \delta(1,3) &=& \sum_2 \mathcal{K}(1,2)\mathcal{K}^{-1}(2,3) \nn
     &=& \, + \, (2/p) \,\cd(1) \mathcal{K}^{-1}(1,3) \nn
     && \, - \sum_2 \mathcal{C}(1) \,(\mathcal{A}^{-1}/4)(1,2) \,\mathcal{C}(2) \,\mathcal{K}^{-1}(2,3).
\label{IntegralEquation}
\end{eqnarray}

\subsection{Determining lapse and shift}

    Since the diffeomorphism constraints are a ($\pi$,q) pair,
further commutation of these constraints produces (almost)
nothing new.  However, the bracket of
$D_2$ with H gives a Laplace-like equation for \Nt.
\begin{eqnarray}
    0 &=& 2 (\mathrm{E}^z_Z) \,\cd(\cd \Nt) + \Nt\{ 2 \,\epsilon_{AB} \,\mathrm{K}^A_x \,\mathrm{K}^B_y \,\Etwo \nn
              & &\quad  +2 \,\mathrm{K}^Z_z(n) \,\mathrm{E}^z_Z \,\mathrm{K}^A_a(n) \,\mathrm{E}^a_A(n) \nn
              & &\quad + (1/2)( \cd \mathrm{E}^y_Y /\mathrm{E}^y_Y  -\cd \mathrm{E}^x_X /\mathrm{E}^x_X)^2 \nn
                &&\quad -  (\cd \Etwo/\Etwo) \,(\cd \mathrm{E}^z_Z ) +(1/2)(\cd \mathrm{E}^z_Z )^2\}/\mathrm{E}^z_Z.
\label{LaplaceForNt}
\end{eqnarray}
The unidirectional constraints, \eq{NaiveUniConstraints},
force every term in this equation to cancel, except $\cd$\Nt terms.
We have rederived $\cd$\Nt = 0, the
first of the unidirectional
plus diffeomorphism constraints, \eq{UniConstraints}.

    Also, the bracket  of $\mathrm{D}_1$ with $\mathrm{H}_z$
gives
\begin{equation}
    \cd \mathrm{N}^z = 0.
\label{NzEqZero}
\end{equation}

\section{Final form of the Hamiltonian}

    Using the unidirectional constraints, one may eliminate KE products
from the Hamiltonian, \eq{H2}. The $\Gamma$ may be replaced
by functions of \Etld, using \eq{GammaEqualsTriads}.
\begin{eqnarray}
    \mathrm{\Nt\Ht +ST}& = & \sum_n (1/\kappa)\{(\Nt/2)\mathrm{E}^z_Z \,(\cd \mathrm{E}^y_Y /\mathrm{E}^y_Y  -\cd \mathrm{E}^x_X /\mathrm{E}^x_X)^2 \nn
        & &\quad + \cd \mathrm{E}^z_Z \,[- \Nt\,(\cd \Etwo)/\Etwo \nn
        &&\quad + \Nt\,\cd \mathrm{E}^z_Z/2\mathrm{E}^z_Z - \cd \Nt] \}
\label{H3}
\end{eqnarray}
On the last line of \eq{H2} if the $\Sigma$ are replaced
by their values in terms of \Etld, then
\begin{align}
  \mbox{last line}&=[ -\Gamma^A_a(n) \,\epsilon^{ZBA}\cd [ \,\mathrm{E}^a_B(n) \Nt] \nn
                    &= -\Gamma^A_a(n) \mathrm{E}^a_B\,\epsilon^{ZBA} \Nt [\cd  \,\mathrm{E}^a_B(n)]/\mathrm{E}^a_B \nn
                    & \qquad  -\Gamma^A_a(n) \,\epsilon^{ZBA} \,\mathrm{E}^a_B(n) \cd \Nt \nn
                    &= -(\Nt/2)\{-[\Gamma_x^Y \E{x}{X} + \Gamma_y^X \E{y}{Y}][-\cd \E{x}{X}/\E{x}{X} + \cd \E{y}{Y}/\E{y}{Y}]\nn
                    & \qquad + [-\Gamma_x^Y \E{x}{X} + \Gamma_y^X \E{y}{Y}][+\cd \E{x}{X}/|\E{x}{X} + \cd \E{y}{Y}/|\E{y}{Y}]\}\nn
                    & \qquad - \cd \E{z}{Z} \cd \Nt.
\label{LastLine}
\end{align}
The final line uses Gauss, \eq{PartsofGaussIdentity}.  The
$\Gamma \times$ E products may be simplified using \eq{GammaEqualsTriads}.

    The above calculations use the original three unidirectional
constraints, \eq{NaiveUniConstraints}, rather than the
constraints which survive diffeomorphism gauge fixing,
\eq{UniConstraints}.  Consequently the diffeomorphism gauge
is not yet imposed.  (This will be done in the
following paper.)

    In \eq{H3} there is an ST on the left, but no ST
on the right.  The $\cd\hat{h}\,\Sigma$
term in the Euclidean Hamiltonian has been integrated by parts,
and the surface term from that
integration by parts cancels the ST.

    A fine point: \Ht + ST is
not a constraint; it is the true
Hamiltonian.  To get the constraint (\Ht only; no ST)
one must undo the integration by parts, which restores the
ST, then discard the ST.  Undoing the integration by parts
changes only the $\cd \Nt$ term in \eq{H3}:
\begin{equation}
    - \cd \mathrm{E}^z_Z \cd \Nt \rta + \Nt \cd(\cd \E{Z}{z}).
\label{ReverseIBP}
\end{equation}

   One can also simplify $\mathrm{H}_z$, again using the unidirectional
constraints to eliminate K.  Then the scalar and vector constraints
are the same, except for a sign.
\[
    \mathrm{H}_z = -\Ht.
\]
The minus sign is reasonable, since $\partial/\partial Z
= -\partial/\partial T$.

    The number of surviving equations now equals
the number of surviving unknowns.
After the single polarization constraints are introduced
and the triad matrix becomes diagonal,
three diagonal triads remain, plus their associated
momenta, plus lapse and shift. The ($\pi$,q) pair
\[
    (\mathrm{K}^Y_y \mathrm{E}^y_Y - \mathrm{K}^X_x \mathrm{E}^x_X,\; \ln (\mathrm{E}^y_Y/\mathrm{E}^X_x))
    \]
represents the physical degree of freedom and must be fixed
using initial conditions.

    One of the remaining ($\pi$,q) pairs
was fixed by the diffeomorphism gauge conditions $\mathrm{D}_i$.
Requiring consistency of those constraints fixes lapse and shift.

    The two constraints, $\mathrm{H}_z$ = -\Ht = 0, have
collapsed to a single constraint; nevertheless, this single
constraint plus a unidirectional constraint are enough to
determine the remaining non-dynamical pair. This pair is the ($\Pi$,Q) introduced
in section \ref{ElinimateConstraint}. $\mathrm{H} =0$ determines
Q, which is essentially \E{Z}{z}; and
$\Pi$ is related to Q by a unidirectional
constraint.

\section{Discussion}

	This paper was written primarily to provide a theory for use in
in the succeeding paper; but  there is something to
be said for this theory as a goal in itself. In principle,
a classical limit derived from an exact theory
could have unacceptable  properties.  One should
ask, what behavior is appropriate in the classical
limit.  Put another way,  the problem should be approached
from the classical side, as well as from the exact side;
and this is what the present paper does.

    The theory constructed here includes
non-local features.  The good
news is that the non-local features which survive in
small sine limit cause no difficulties.  If the non-local
exact theory treats nearest neighbors symmetrically, then
the Euclidean Hamiltonian will contain the central difference
of a holonomy.  If terms linear in extrinsic curvature
are required to vanish from the LQG Hamitonian, as they do
from the FT Hamiltonian,
then every derivative in the spin connection must be
replaced by a central difference, leading to a
theory with central differences everywhere.  Similarly, the other non-local
feature, the z holonomies, conceivably
could cause problems with commutators, but the commutators are reasonable
because of the slow variation assumption.

    The bad news is that
other non-local features disappear in this limit.
Small sine calculations are not a good way to distinguish between
various non-local versions, since they all tend to possess
the same, universal, small sine limit.  The non-local
example discussed in section \ref{Exact} uses forward differences
and unaveraged z holonomies, whereas the SS limit contains
central differences and the averaged holonomies $\hat{h}_z(n_z)$
defined at \eq{Commutator2}.  Similarly, the non-local theory
constructed in \ref{nonlocal}
starts from "single grasp" \Etld operators, i.~e.
operators which grasp only ingoing holonomies at the vertex,
or only outgoing holonomies.
In the SS limit, the single grasp \Etld disappear,
replaced by \Etld which grasp both holonomies.

     The calculation required various identities involving the $\Gamma$.  Those
identities hold in FT, and also in semiclassical LQG,
because the proofs
use algebra, rather than calculus or properties of the derivative.
The  identities therefore continue to hold even after
derivatives are replaced by differences.

    In the small sine limit, it is easy to separate
extrinsic curvature from spin connection.  Although the Thiemann
procedure generates only the combination $\gamma \rK = \rA - \Gamma$, rather
than $\Gamma$, the
symmetries force off-diagonal $\rA^A_b$ and on-diagonal
$\Gamma^A_a$ to vanish.  This circumstance allows us to
separate out the $\Gamma$.
On-diagonal K's are pure extrinsic curvature,
while off-diagonal "K's"  are pure $\Gamma$'s.

    It may be possible to separate out the $\Gamma$,
even when higher powers of sine are included.
Note the $\Gamma$ are odd under n+1 $\leftrightarrow$ n-1
(because they contain a central difference),
while the K's are even under this interchange, and
contain one higher power of sine.
If this pattern persists to higher orders in sine, it
should be possible to split off the spin connection (odd) part of
the K tensor, and check whether the SS identities
continue to hold in higher order.

    Readers who are familiar with the relation between
geometrodynamical variables, Szekeres
variables, and
(K,\Etld) variables, will recognize numerous points
where the theory shifts to combinations of the K and \Etld
which equal Szekeres or geometrodynamical variables \cite{Szekeres}.
For example, the ADM $\pi^{ij}$ are linear
combinations of K
$\times$ \Etld products.  The triad combinations
involving logarithms are Szekeres variables.
Whatever the superiority of K and \Etld at short
distances, the traditional combinations hold the edge
in the SS limit.

    Of course K $\cdot$ E is
not really a geometrodynamical variable, because the
"K" is a holonomy, not a field.  This is the fundamental
change which leads to quantization of areas and volumes.
Nevertheless, the combination holonomy times triad seems to
be more appropriate than holonomy alone.

    In this paper, the small sine approximation
was used to simplify the Hamiltonian.
Suppose, however, one retains the small sine assumption,
even near e = 0.   (Near e = 0 one must
abandon slow variation, of course, and regulate the cotriads.)
Then one has a small sine \emph{model} which nevertheless retains the
most desirable features of full LQG:
geometrical quantities are quantized, the connection
remains within a bounded holonomy, and the model
has a simpler Hamiltonian.

    The small sine model is especially convenient in the
plane wave case. Given the shift from N to \Nt,
\[
    \rN \mathrm{E}^j_J \,\mathrm{E}^k_K \,\epsilon^{IJK}/\mid e \mid = \Nt \,\mathrm{E}^j_J \,\mathrm{E}^k_K \,\epsilon^{IJK}/\mathrm{E}^z_Z,
\]
one needs to regulate only
$1/\mathrm{E}^z_Z$.  Bannerjee and Date replace
the 1/\E{z}{Z} by two factors of
\begin{eqnarray}
    &(8/\kappa\gamma)h_z \,[\,h_z(n)^{-1},\sqrt{\si(z) \mathrm{E}^z_Z(n)}\,]  \quad \mbox{(LQG)} \nn
    &\rightarrow \mbox{\si(z)} \,\sigma_Z /\sqrt{\si(z) \mathrm{E}^z_Z}. \quad \mbox{(QFT)}
\label{RegularizationSqrtEz}
\end{eqnarray}
sign(z) is the sign of \E{z}{Z}.  For a similar
maneuver in a cosmological context, see Bojowald \cite{Bojowald2}.
Because  $\mathrm{E}^z_Z$ is already diagonal,
its operator square root is immediate.

    In a SS model, the spin connections also need to be
regulated.  From CGR formulas
for $\Gamma \cdot$E, \eq{SSKGamma}, with derivatives
replaced by differences, plus CGR formulas for the
cotriads, we get
\begin{eqnarray}
    \Gamma^X_y \mathrm{E}^y_Y + \Gamma^Y_x\mathrm{E}^x_X &=& \si(e) \, [-(\cd e^Y_y) e^X_x + e^Y_y \cd e^X_x] \nn
    &=& \si(e) \,(\mathrm{E}^z_Z/\mid e \mid)^2 \,[ \,(\cd \,\mathrm{E}^y_Y) \mathrm{E}^x_X - \mathrm{E}^y_Y (\cd \mathrm{E}^x_X) \,].
\label{RegulatedGamma}
\end{eqnarray}
The dangerous overall factor of $(\mathrm{E}^z_Z/\mid e \mid)^2$
can be removed by an appropriate choice of gauge.
For p = 1/2, that factor becomes a constant.
The other linear combination follows from Gauss and
is  free of singularities.
\[
    \cd \mathrm{E}^z_Z = -\Gamma^X_y \mathrm{E}^y_Y + \Gamma^Y_x\mathrm{E}^x_X.
\]

    One could argue the small sine model should always be the
first model tried, when testing LQG.  Small sine keeps the order
($\sin$ + $\sin^2$) terms in the constraints.  These are the
only terms we are sure of, because they supply the correct
FT limit.

\appendix

\section{A Non-local model}
\label{nonlocal}

    This appendix constructs an
exact LQG Euclidean Hamiltonian.   It starts from field
strengths which are nearest neighbor non-local,
and z holonomies integrated from vertex to vertex,
like the non-local model discussed in
the main body of the paper.
However, the \Etld are "single grasp"; they grasp at only
one of the six surfaces of the cube surrounding
each vertex.   The main result of this
appendix is: in the semiclassical
limit, single grasp \Etld are replaced
by double grasp, \Etld which grasp both the incoming and outgoing
holonomy at each vertex.

    This appendix uses the "congruence" rather
than $\rS_1$ picture for the topology of the spin
network.  (Topology is discussed at the beginning
of section \ref{Approx}.)  The congruence picture
is better for determining the number
of loops which contribute to each exact
field strength.

    Recall the quantum field theory expression for the
Euclidean Hamiltonian.
\begin{eqnarray}
   - \mathrm{N H}_e + \mathrm{ST} &=&   \int d^3x \mathrm{N (F}^I_{jk} \,\mathrm{E}^j_J \,\mathrm{E}^k_K \,\epsilon_{IJK}/2\kappa \mid e \mid) + \mathrm{ST}; \nonumber \\
   \mathrm{ST} &=& -(\mathrm{N \,A}^I_a \mathrm{E}^z_Z \mathrm{E}^a_K\epsilon_{IZK}/\kappa \mid e \mid)\mid_{z=-\infty}^{+\infty} \:;\nn
    \kappa & =&  8 \pi \mathrm{G}; \nn
     |e|  &:=&  \sqrt{\mbox{det} \,\mathrm{E} \, \si(e)}.
\label{ClassicalHe2}
\end{eqnarray}

    The LQG formula for field
strengths in \eq{ClassicalHe2} generalizes the classical formula
\begin{equation}
        \mathrm{F}_{ij} = \mbox{lim}_{\Delta A\rta 0} (\prod h)/\Delta A(ij),
\label{FijEqhLoop}
\end{equation}
where $\prod h$ is the product of holonomies around
the edges of infinitesimal area $\Delta$ A(ij).  A local
treatment of the Hamiltonian would construct each F using small areas
at a single vertex $n_z$, plus triads at that vertex; then sum over vertices.
I.e. the basic modular unit would be the vertex.

    However, for a non-local Hamiltonian, one must use non-local
modular units.  Since \eq{FijEqhLoop} contains an area, the appropriate
basic modular units should be areas.  The areas are bounded by
nearest neighbor vertices:  i.e. holonomies along
each edge of the area run from one vertex to the next,
nearest neighbor vertex.  For example, the LQG contribution to
$\mathrm{F}_{xy}$ from the area bounded by vertices $(n_x,n_y,n_z),
(n_x,n_y+1,n_z),(n_x+1,n_y+1,n_z),(n_x+1,n_y,n_z)$ is
\begin{equation}
    \mathrm{F}_{xy} = \rmi \,\mbox{const.} \,h_y^{-1}h_x^{-1}h_x h_y + \rH.c. \quad \mbox{(LQG)}
\label{LQGFxy}
\end{equation}
Each $h_i$ traverses edge i from $n_i$ to $n_i + 1$; the $h^{-1}$ traverse in
the reverse direction.  Reading from right to left, the explicitly
written term circulates the xy area in counterclockwise direction.
The Hermitean conjugate (H.c.) term circulates in the clockwise direction.

    The above two terms are not the only possibilities, even though we restrict ourselves to
circuits starting from $(n_x,n_y, n_z)$ and continuing in the xy plane to nearest neighbors only.
In fact there are eight such terms.  The eight
correspond to the four vertices in the xy plane which are nearest
neighbors to the vertex $(n_x,n_y, n_z)$, times two for clockwise or
counterclockwise circuit.  The eight may be grouped into
four sets of two terms each,
after we impose the requirement that the field strength is Hermitean.  The full LQG
expression for $\mathrm{F}_{xy}$ therefore contains four adjustable constants
analogous to the "const" in \eq{LQGFxy}.  One could
determine them by carrying out a small sine expansion, and demanding that the
expansion have the minimum number of powers of sine beyond those needed to recover the
quadratic limit.  The discussion would be straightforward but lengthy, and
will be omitted in order to focus on the semiclassical limit.

    One can small-sine expand
the contribution \eq{LQGFxy}, using \eq{SSLimitHhat}.
Keeping up to order $(\sin)^2$, one gets
\begin{equation}
    \mathrm{F}_{xy}(n_z) = 2\,\rmi [ \,\hat{h}_x(n_z),\hat{h}_y(n_z) \,], \:\mbox{(SS)}
\label{SSFxy}
\end{equation}
All eight loops give the same small sine limit.
The $\hat{h}_a$ need only a single argument $n_z$, since
all variables are independent of x and y.

    \Eq{SSFxy} has no linear-in-sine terms.  These would spoil
the QFT limit
\begin{eqnarray}
    \mathrm{F}_{ij}  &\rightarrow&  (\partial_i \mathrm{A}^I_j - \partial_j \mathrm{A}^I_i \nn
            & & + \mathrm{A}^I_i \mathrm{A}^J_j \epsilon_{IJK}) \,\sigma^I \Delta x^i \Delta x^j. \:\mbox{(QFT)}
\label{QFTFij}
\end{eqnarray}
For ij = xy, the linear in A terms are absent, because all fields are
independent of x and y.

    The remaining two field strengths, $\mathrm{F}_{za}$ with a = x,y, may be constructed
in similar fashion.  For $\mathrm{F}_{zx}$, for example, the relevant plane is zx
rather than xy; but there are still eight loops, corresponding to four vertices which are nearest
neighbors to the vertex $(n_x,n_y, n_z)$, times two, for clockwise or anticlockwise.
Four of the holonomy loops travel forward to  $n_z+1$;
the remaining four travel backward to  $n_z-1$.
The four "forward" loops are
\begin{align}
   \rF_{za}(n_z,n_z+1) &= \rmi \,\mbox{const.}\, [h_z(n_z,n_z+1)^{-1}h_a^{-1}(n_z+1)h_z(n_z,n_z+1) h_a(n_z) \nn
                   &\qquad   - (h_a\leftrightarrow h_a^{-1})+ \rH.c \,]. \qquad \mbox{(LQG)}
\label{LQGFza}
\end{align}
The $(h_a\leftrightarrow h_a^{-1})$ term is the second forward loop;
"H.c." includes the remaining two loops, which have clockwise $\rla$
anticlockwise.  Since the $h_z$ link two vertices, $ \mathrm{F}_{za}$
requires two $n_z$ arguments.

    In the xy case all four areas had the same small sine limit. In the za case,
forward and rearward loops have different SS limits and cannot be
combined, because they involve different x holonomies $h_x(n_z\pm 1)$ on
different x edges.

    The first two forward loops in
\eq{LQGFza} each contain unwanted
${h_x(n_z) \,h_x(n_z+1)}$ quadratic terms in the SS limit; the minus sign
between these two loops insures that the unwanted terms cancel.  Explicitly,
the small sine limits are
\begin{align}
   2 \,\mathrm{F}_{za}(n_z,n_z+1)
            &\rightarrow (-2 i)\{-[ \,\hat{h}_z(n_z,n_z+1),\hat{h}_a(n_z+1) \,] \nn
            &\quad +\hat{h}_a(n_z+1) - \hat{h}_a(n_z) \nn
             &\quad    +[ \,\hat{h}_a(n_z+1),\hat{h}_a(n_z) \,]\} - (\hat{h}_a \leftrightarrow -\hat{h}_a); \nn
    2 \,\mathrm{F}_{za}(n_z,n_z-1)
            &\rightarrow (-2 i)\{-[ \,\hat{h}_z(n_z-1,n_z),\hat{h}_a(n_z-1) \,] \nn
            &\quad +\hat{h}_a(n_z) - \hat{h}_a(n_z-1) \nn
            &\quad + [ \,\hat{h}_a(n_z-1),\hat{h}_a(n_z)] \,\} - (\hat{h}_a \leftrightarrow -\hat{h}_a). \quad  \mbox{(SS)}
\label{SSFza}
\end{align}
The $\hat{h}\leftrightarrow -\hat{h}$ terms represent the second loop; the unwanted
$(\hat{h}_a)^2$ terms cancel out when the two loops are summed.

    The classical xy field strength distorts a spherical cloud of test
particles into an ellipsoid.  This distortion mimics the behavior of
the sphere under free fall in a \emph{static} gravitational field.
The za field strengths produce the distortions typical of
time-varying waves.  It is not surprising,
therefore, that the za field strengths possess all the
nonlocality.

    Now consider the triads.  Since triads are associated with areas and volumes,
which are local, I assume the LQG triads are local, with support
at the vertices. More precisely: draw a small cube around each
vertex.  The triads, which are associated with area two forms
via \E{a}{A}$ \, \epsilon_{abc} \, dx^b\wedge dx^c$, live on the six
faces of the cube.

    When defining the volume operator, typically
one assumes each \E{a}{A} grasps at both faces having outward normals
in the "a" direction, the positive face (with normal pointing in the
positive a direction) and the negative face.  This is natural;
the volume operator is not directional, and one would expect
contributions from all six faces.  The volume operator therefore involves
the product of three operators, each grasping at two faces.
\begin{equation}
    \mathrm{E}^i_I := \mathrm{E}^i_I(+) + \mathrm{E}^i_I(-),
\label{defEpm}
\end{equation}
where  $\pm$, the sign of the normal, indicates where
the \Etld grasps.  However, if one considers \Etld other
than those involved in the volume operator, the double
grasp \Etld is not especially natural.

    For example, in the operator expression for Gauss'
law, one needs the \emph{difference} between
the + and - operators, \E{z}{Z}(+) - \E{z}{Z}(-),
because Gauss' Law involves the difference between
ingoing and outgoing Z component of spin.  It is not
enough, therefore, to specify the vertex $n_z$ where
the triad has its support; one must also supply  an
argument $\pm$ specifying which face it grasps.

    In \eq{ClassicalHe2}  there are two triads and a
volume e associated with each field strength.  A given
field strength is a holonomy loop passing through
four vertices, but at each vertex, the holonomies do not
pass through all six faces.  One holonomy enters at one face, and
a different holonomy leaves at another face.   Therefore one
can associate a face with each triad as follows.
(If a given loop on area ij starts and ends at a vertex $n_z$, then
the triads associated with that loop are all evaluated at
$n_z$; and) if a holonomy $h_i$ at $n_z$ passes through a face
having sign + (-), then the
triad \E{i}{A} is given the sign + (-).

    For example, consider the xy holonomy loop beginning at vertex
($n_x,n_y,n_z$) and passing through the
+ x face to ($n_x+1,n_y,n_z$).  The loop
then continues to ($n_x+1,n_y+1,n_z),
\cdots$ , finally returning down the y axis through
the + y face.  Both holonomies pass through + faces; the
triads associated with this contribution
would be \E{x}{A}(+) and \E{y}{A}(+).   The same
triads occur in the H.c. loop.

    The remaining three pairs of loops involve the remaining three
sign pairs: ($\pm,\mp$) and (-,-).  The $\mathrm{F}_{xy}$ contribution
remains a sum of four terms, if we group each term and its
H.c. together as a single contribution.  Now each term contains
a different sign pair.

    A z triad \E{z}{Z} also contributes to each $\mathrm{F}_{xy}$
term in $\mathrm{H}_e$.  The z triad is contained in the factor of
volume, e.  It is not obvious what sign to assign to the z
triad, since no xy holonomy
passes through a z face.  However, a given xy area
bounds two volumes, each containing a different z line ($n_z \pm 1,n_z$).
One can get either sign, $\pm$ 1, according as
the xy area is interpreted as bounding the volume containing ($n_z,n_z +1$),
or ($n_z - 1,n_z$).  There is no reason
to favor one interpretation over the other, and we therefore make the
z triad the sum of the two z signs: \E{z}{Z} = \E{z}{Z}(+) + \E{z}{Z}(-).

    In the small sine,
slow variation limit, the triad functions in $\mathrm{H}_e$ lose their
dependence on the individual \E{i}{I}($\pm$) and depend only on  the
sums \E{i}{I} := \E{i}{I}(+) + \E{i}{I}(-).
Proof: begin with the four loops contributing to $\mathrm{F}_{xy}$.
All field strengths have the same small sine limit, \eq{SSFxy}.
One can factor this out, leaving a sum over four triad functions.
\begin{equation}
    \mathrm{H}_e(xy) = \mathrm{F}_{xy}\mathrm{(SS)} \sum_{\eta }f[ \,\rE^x(\eta_x), \,\rE^y(\eta_y), \,e(\eta_x, \eta_y) \,]\nn
\label{xyTriadSeries}
\end{equation}
where $\eta_a = \pm 1)$ indicate the areas grasped by the \Etld;
and the \E{z}{Z}, not indicated explicitly, are already in the
desired form.

    Now expand each transverse \Etld:
\begin{eqnarray}
    &\mathrm{E}^a(\pm) &= [(\mathrm{E}(+) + \mathrm{E}(-)]^a/2 \pm [(\mathrm{E}(+) - \mathrm{E}(-)]^a/2 \nn
            &&= \mathrm{E}^a/2 \pm \td \mathrm{E}^a/2 \nn
            &&= (\mathrm{E}^a/2)[1 \pm \td \mathrm{E}^a/\mathrm{E}^a].\nn
    &\td \mathrm{E}^a(n) &:= [(\mathrm{E}^a(+) - \mathrm{E}^a(-)].
\label{defTildeDifference}
\end{eqnarray}
The usual difference $\delta f$ denotes the difference between values
of f evaluated at two different vertices.  The tilde difference denotes
the difference between two values of f located at the same vertex,
but on opposite faces of the vertex. I
insert the expansions of \eq{defTildeDifference} into \eq{xyTriadSeries},
and power series expand around $\tilde{\delta}\mathrm{E}=0$,  assuming the
tilde differences are small because of the slow variation assumption.

    The expansion contains no
terms having an odd number of tilde differences, since the sum in
\eq{xyTriadSeries} is even under (+ $\leftrightarrow$ -).  The
expansion of an arbitrary symmetric function of the
\E{a}{A}($\pm$) begins with the terms
\[
    f[ \,\mathrm{E}(+) \,] + f[ \,\mathrm{E}(-) \,] = 2 \, f(\mathrm{E}/2) + (1/2!)\,(\partial^2 f/\partial \mathrm{E}^2) \, (\td \mathrm{E})^2.
\]
The leading term in the expansion contains two more factors
of $\mathrm{E}$ than the second order term, and is therefore
larger than the second order term by a factor
\[
    (\tilde{\delta}\mathrm{E})^2/(\mathrm{E}^a)^2 = \mbox{order}(\delta f/f)^2 .
\]
The second order term can be neglected.    Tilde
differences have disappeared from the $\mathrm{F}_{xy}$ terms.

    Now consider the $\mathrm{F}_{za}$ terms, for example the forward
areas involving z holonomies on edge $(n_z, n_z+1)$, and za = zx.
At \eq{SSFza} the sign between two terms was adjusted so as
to cancel an unwanted commutator term.  This cancelation must be
reconsidered; the two terms are now multiplied by different
triads $\mathrm{E}^x(\pm)$.  The unwanted brackets now have
a contribution of the form
\begin{eqnarray}
    [\,\hat{h}_x(n_z+1),\hat{h}_x(n_z)\,]&\times(f[ \,\mathrm{E}^x(+),\mathrm{E}^z(+) \,] - f[ \,\mathrm{E}^x(-),\mathrm{E}^z(+) \,]) \nn
                        &\cong [\,\delta_f\hat{h}_x(n_z), \,\hat{h}(n_z)\,] 2 \, (\partial f/\partial \mathrm{E}^x) \, \tilde{\delta}\mathrm{E}^x .
 \end{eqnarray}
This term is second order in differences and can be dropped.

    The remaining terms, those without the unwanted commutators,
give the correct QFT limit and are even under
$\mathrm{E}^a(+)\leftrightarrow\mathrm{E}^a(-)$.  By the same argument as for
the $\mathrm{F}_{xy}$ terms, the expansion in powers of $\tilde{\rE}^a$
may be terminated at the leading term which is independent of  $\tilde{\rE}^x$.
Similarly for the rearward loops.

    At this point the $\rF_{za}$ loops have the desired $\rE^a$ dependence, but forward
loops are multiplied by a function of $\mathrm{E}^z(+)$, while the
rearward loops depend on $\mathrm{E}^z(-)$.  From the SS limits, \eq{SSFza},
both loops contain forward difference
terms $\delta_f \hat{h}_a$ and commutator terms $[\hat{h}_z,\hat{h}_a]$.

    Consider first the difference terms.  As at \eq{defTildeDifference},
expand in sums plus differences and drop the term linear in differences.
\begin{align}
   \delta_f h_a(n_z) f[\mathrm{E}^z(+)] + \delta_f h_a(n_z&-1) f[\mathrm{E}^z(-)]\nn
    &\cong [\delta_f h_a(n_z)+\delta_f h_a(n_z-1)]f[\mathrm{E}^z] \nn
    & \quad + [\delta_f h_a(n_z)- \delta_f h_a(n_z-1)](\partial f/\partial\mathrm{E}^z )\td \mathrm{E}^z.
\end{align}
The bracket on the middle line is twice $h_z(n_z)$, from definitions
\eqs{defFirstDifference}{defSecondDifference}, and slow variation.  The last line is
down by two factors of $\delta f/f$ and may be dropped: the square bracket on the last line
is the second forward difference of h.

    For the commutator terms one must  expand, not only
f[$ \,\mathrm{E}^z(+) \,$], but also $\hat{h}_z$ and $\hat{h}_a$.
\begin{align}
    \hat{h}_z(n_z,n_z \pm1)
   & = [\hat{h}_z(n_z,n_z+1)+\hat{h}_z(n_z-1,n_z)]/2 \pm \td \hat{h}_z(n_z-1,n_z)/2 \nn
    & := \hat{h}_z(n_z) \pm \td \hat{h}_z(n_z-1,n_z)/2; \nn
   \hat{h}_a(n_z \pm 1) &= \hat{h}_a(n_z) \pm \delta_f \hat{h}_a(n_z\pm1).
\label{defhzpm}
\end{align}
The commutator terms are then
\begin{eqnarray}
  \sum_{\pm}[\hat{h}_z(n_z,n_z \pm 1)&,&\hat{h}_a(n_z \pm 1)] \,f[ \,\mathrm{E}^z(\pm) \,] \nn
    &=& [\hat{h}_z(n_z),\hat{h}_a(n_z)] \, 2 \, f(\mathrm{E}^z) + \mbox{order} (\cd f/f)^2.
\end{eqnarray}
The two terms linear in $\td \hat{h}_z(n_z-1,n_z)$ and $\td \mathrm{E}^z$
have opposite signs and cancel.  The two terms linear in $\delta_f \hat{h}_a$
have the form
\[
    [\hat{h}_z(n_z),\delta_f \hat{h}_a(n_z) - \delta_f \hat{h}_a(n_z-1)] \,f[ \,\mathrm{E}^z \,].
\]
This difference of differences is a second forward difference,
which is order $(\delta f/f)^2$.

    All four $\mathrm{F}_{za}$ loops now depend only on $\mathrm{E}^i = \mathrm{E}^i(+) + \mathrm{E}^i(-)$.
Dependence on the $\mathrm{E}^i(\pm)$ has disappeared.  $\Box$

\section{Volume and area ambiguities}

    It is necessary
to take the square root of \E{z}{Z} when \E{z}{Z} is a
factor in the volume squared operator; or when factors of
$1/\sqrt{\rE^z_Z}$ must be regulated in the Hamiltonian.
    There is an ambiguity in the definition of the volume operator,
and this ambiguity is inherited by the area operator \E{z}{Z}.
When two z holonomies
terminate at a vertex, the \E{z}{Z}  operator generates two amplitudes.
To parallel a  terminology from classical optics: should one add
first, then take the square root?  Or take the square root first,
then add?  I.e. should one
add the amplitudes from each z holonomy, then take
the square root of
the magnitude of the result; or should one take the square
root of the magnitude of each amplitude first, then add?
let \E{z}{Z} act on two eigenfunctions with eigenvalues
($m_i,m_f$) or two coherent states with peak values ($m_i,m_f$).
\begin{align}
   \sqrt{\mid \rE^z_Z \mid} \ket{m_f}\ket{m_i} &= \sqrt{\mid m_f + m_i \mid/2}\ket{m_f}\ket{m_i}; \quad \mbox{(add first)}\nn
   \sqrt{\mid \rE^z_Z \mid} \ket{m_f}\ket{m_i} &= (\sqrt{\mid m_f \mid/2} + \sqrt{\mid m_i \mid/2})\ket{m_f}\ket{m_i}. \, \mbox{(square root first)}
\end{align}
The factors of 1/2 come from the integrations over half a delta function.

    In the case of the volume operator, the literature contains advocates for both
"add first" \cite{ALvolumeI,ALvolumeII} and "take the square root first"
\cite{RSvolume,DePRvolume} choices.  For a
discussion of the distinct regularization schemes leading to each
choice see reference \cite{ALvolumeII}.

    This paper adopts the ''add first" choice, for the following
(non-rigorous) reason. In the special case $m_f = m_i$, it is
possible to view these
two eigenfunctions ($\ket{m_f}\ket{m_i}$) as a single
eigenfunction at the vertex.  When this is grasped by
the  \E{z}{Z}, the contribution equals
$m_i$ with no 1/2.   This result for one eigenfunction equals
the $m_f = m_i$ limit of the result for two different eigenfunctions,
only if the "add first" prescription is used.

\section{The surface term}
\label{surface}

    The $\cd (\hat{h}_a) \,e^{za}$
terms in $\mathrm{H}_e$ contain a second derivative, since
$\hat{h}_a$, when expressed in terms of tetrads, contains a
time derivative.  The $\cd$ must be integrated by
parts (IBP) onto the cotriad.  The IBP brings the Hamiltonian
(and Lagrangian) into a standard form with only first derivatives.
The  IBP generates a
total derivative, which becomes a surface term $\hat{h}_a  \, e^{za}$.
This term must be canceled, which means a ST must be added to
perform the cancelation:
If
\begin{eqnarray*}
    \mathrm{H}_e \sim (\cd \hat{h}_a) \,e^{za}+\cdots &=& \cd (\hat{h}_a e^{za})- \hat{h}_a  \,\cd e^{za} + \cdots\\
                                                &:=& -\mathrm{ST} - \hat{h}_a  \,\cd e^{za} + \cdots,
\end{eqnarray*}
then
\[
    \mathrm{H}_e + \mathrm{ST} \sim -\hat{h}_a  \,\cd e^{za} + \cdots; \:\mbox{and no ST}.
\]
The $\cdots$ denote terms
which are derivative-free and do not contribute to the ST.
The integration by parts shifts the difference onto the
triads and changes the sign of a term in the Hamiltonian.

    Although the surface term produces  only a
rather simple change in $\mathrm{H}_e$, the ST must be
calculated in detail, because
ST is the physical Hamiltonian.  $\mathrm{H}_e$
is a constraint, and vanishes when acting on physical
states.  The ST does not vanish.
In the follow-on paper, the ST is used to
compute the total energy of the solution constructed in that paper.

    From the preceding discussion, the surface term comes entirely
from the $\cd \hat{h}$ terms in $\mathrm{F}_{za}$.
Insert \eq{SSHe} for
$\mathrm{F}_{za}$ into \eq{HeWithNt} for $\Ht_e$:

\begin{align}
  - \Nt \,\Ht_e + \mathrm{ST} &=  (1/2\kappa )\sum_n \{ \cdots   + (-2i)(\delta_c \, \hat{h}^C_a) \mathrm{E}^a_B \,\epsilon^{ZBC} \,\Nt(n) \} +\mathrm{ST} \nn
       &= (1/2\kappa )\sum_n \{ \cdots - (-2i)\hat{h}^C_a \delta_c \,  [ \,\mathrm{E}^a_B \,\epsilon^{ZBC} \,\Nt(n) \,] \nn
       &  \quad + (-2i)\delta_c \, [ \,\hat{h}^C_a \,\mathrm{E}^a_B \,\epsilon^{ZBC} \,\Nt(n) \,] \} +\mathrm{ST}.
\label{LQGHeDifferenceTerms}
\end{align}
\Eq{ApproximateDistribution} was used to carry out the difference analog of
integration by parts.

    This is a good point to describe the labeling of
the vertices at the surface.  The  $\sum_n$ in the Hamiltonian
ranges from n = min to
n =  mx; min $\leq$ n $\leq$ mx.  However, the spin network
itself extends to values  n $<$ min and n $>$ mx.  Compare
classical field theory, where one integrates
the Lagrangian or Hamiltonian from min z to mx z, but the space
extends beyond these limits.

    In principle, the limits (min, mx) can be chosen anywhere.  In practice,
the limits are chosen to lie in an asymptotic region, so that surface terms generated
by integration by parts can be evaluated using boundary conditions.
Similarly here, the only restriction on min and mx is that the system
is asymptotic at those values of n; but the spin network does not vanish
beyond those limits.

    In particular, suppose the construction of the
spin connection predicts that it depends on a central difference
\[
[ \,f(n+1) - f(n-1) \,]/2.
\]
 At n = mx, f(n+1) is f(mx +1).
This quantity is not assumed to vanish.

     Since the $\cd$ connects every other
vertex, the total derivative on the last line, \eq{LQGHeDifferenceTerms},
gives rise to
two surface terms, one from even n terms and one from odd
n.  $\Ht_e$ becomes

\begin{align}
  - \kappa (\Nt \Ht_e + \mathrm{ST})  &= \sum_n\{ \cdots  - (-2i) \hat{h}^C_a \,\delta_c \,[\,\mathrm{E}^a_B \,\epsilon^{ZBC} \,\Nt\,](n) \}  \nn
        & \quad + [\,\Nt(n) (-2i)\hat{h}^C_a \,\mathrm{E}^a_B \,\epsilon^{ZBC}\,](n)[\,\mid_{n = min}^{mx}+ \mid_{n = min -1}^{mx+1}\,](1/2) +\mathrm{ST}.
\label{HeSmallSine}
\end{align}
The ST is now chosen so that the last line vanishes.

    The 1/2 in the surface term comes from the 1/2 in the
central difference.  For example, use the definition of
central difference, \eq{defFirstDifference}, to expand each term in the sum
\[
    \sum_{-1}^{+1} \cd f(n) = (1/2)[\,f(+2) - f(-2) + f(+1) - f(-1)\,].
\]

    It is possible to eliminate the holonomy from the surface term.
Replace the $(-2i)\hat{h}^C_a$ by $(\gamma \,\rK + \Gamma)^C_a$ (\eq{SSK}).
The term involving K,
\[
    \rK^C_a \,\mathrm{E}^a_B \,\epsilon^{ZBC},
\]
is (one half of) the Gauss constraint, \eq{PartsofGaussIdentity}, and may be
dropped.  The term involving $\Gamma$ may be simplified by using the
other half of the constraint.
\[
    \Gamma^C_a \, \rE^a_B \epsilon^{BC} = \cd \rE^z_Z.
\]
The surface term is then
\begin{equation}
    \mathrm{ST} = -\,\Nt \, \cd \rE^z_Z(n)\,[\mid_{n = min}^{mx}+ \mid_{n = min -1}^{mx+1}](1/2 \kappa).
\label{EzST}
\end{equation}

\section{Number of vertices}

    This calculation uses a lattice with a fixed number of
vertices. Where does the number of vertices enter into
the calculation?  To obtain the classical limit, I must assume the
fields vary slowly from vertex to vertex, so that the discrete
structure of the spin network is not obvious. The
precise value of the number of vertices is not
important, provided the number of vertices is large
enough to guarantee slow variation.  One could replace
the fixed number of vertices with a distribution in the
number of vertices and nothing would change, provided the
distribution were peaked at a large number.

    The restriction to a fixed number of vertices
may be more apparent than real, because
the classical limit uses coherent states.
In a coherent state, at each vertex n,
the values of SU(2) angular momentum L
are Gaussian distributed.  This distribution
includes angular momentum zero.  In that
sense a coherent state already includes the
possibility of no vertex at n.

    Since the distribution is Gaussian, the
probability of no vertex is very small.
Presumably spin networks
with small numbers of vertices do not
contribute significantly in the classical limit.
\bigskip


\end{document}